\documentclass[sigconf,preprint]{acmart}

\AtBeginDocument{%
  \providecommand\BibTeX{{%
    \normalfont B\kern-0.5em{\scshape i\kern-0.25em b}\kern-0.8em\TeX}}}

\setcopyright{acmcopyright}
\acmPrice{15.00}
\acmDOI{10.1145/3395363.3397372}
\acmYear{2020}
\copyrightyear{2020}
\acmSubmissionID{issta20main-id31-p}
\acmISBN{978-1-4503-8008-9/20/07}
\acmConference[ISSTA '20]{Proceedings of the 29th ACM SIGSOFT International Symposium on Software Testing and Analysis}{July 18--22, 2020}{Virtual Event, USA}
\acmBooktitle{Proceedings of the 29th ACM SIGSOFT International Symposium on Software Testing and Analysis (ISSTA '20), July 18--22, 2020, Virtual Event, USA}

\usepackage{microtype} 

\usepackage{amssymb}
\usepackage{hyperref}
\usepackage{caption}
\usepackage{graphicx}
\usepackage{xcolor}
\usepackage{float}
\usepackage{listings}
\usepackage[T1]{fontenc}
\usepackage{amsmath,amsfonts,amsthm}
\usepackage{algorithmic}
\usepackage{xspace}
\usepackage{makecell}
\usepackage{pifont}
\usepackage{setspace}
\usepackage{adjustbox}
\usepackage{algorithm2e} 
\usepackage{bold-extra}
\usepackage{subcaption}
\usepackage{enumitem}
\usepackage{balance}
\usepackage{bold-extra}
\usepackage[normalem]{ulem}

\usepackage{framed}
\definecolor{shadecolor}{gray}{0.9}
\newcommand{\boxedspit}[1]{
\begin{shaded*}
\vspace{-1mm}
\noindent{\bf\small Example.} #1
\vspace{-1mm}
\end{shaded*}
}

\newcommand{\nextnr}{\stepcounter{AlgoLine}\ShowLn}
\makeatletter
\newcommand{\removelatexerror}{\let\@latex@error\@gobble}
\makeatother

\newcommand{\cmark}{\ding{51}}%
\newcommand{\xmark}{\ding{55}}%

\newcommand{\ourname}{{\sc {Weizz}}\xspace}
\newcommand{\weizz}{{\sc {Weizz}}\xspace}
\newcommand{\afl}{{\sc {AFL}}\xspace}
\newcommand{\driller}{{\sc {Driller}}\xspace}
\newcommand{\qsym}{{\sc {Qsym}}\xspace}
\newcommand{\aflpp}{{\sc {AFL++}}\xspace}
\newcommand{\steelix}{{\sc {Steelix}}\xspace}

\newcommand{\hongfuzz}{{\sc {HonggFuzz}}\xspace}
\newcommand{\vuzzer}{{\sc {Vuzzer}}\xspace}
\newcommand{\redqueen}{{\sc {RedQueen}}\xspace}
\newcommand{\slf}{{\sc {SLF}}\xspace}
\newcommand{\angora}{{\sc {Angora}}\xspace}
\newcommand{\eclipser}{{\sc {Eclipser}}\xspace}
\newcommand{\taintscope}{{\sc {TaintScope}}\xspace}
\newcommand{\tfuzz}{{\sc {T-Fuzz}}\xspace}
\newcommand{\spike}{{\sc {Spike}}\xspace}
\newcommand{\peach}{{\sc {Peach}}\xspace}
\newcommand{\aflsmart}{{\sc {AFLSmart}}\xspace}

\newcommand{\langfuzz}{{\sc {LangFuzz}}\xspace}
\newcommand{\superion}{{\sc {Superion}}\xspace}
\newcommand{\nautilus}{{\sc {Nautilus}}\xspace}
\newcommand{\grimoire}{{\sc {Grimoire}}\xspace}

\newcommand{\lafintel}{{\sc {LAF-Intel}}\xspace}
\newcommand{\klee}{{\sc {KLEE}}\xspace}

\newcommand{\qemu}{{\sc {QEMU}}\xspace}

\newcommand{\firststage}{surgical\xspace}
\newcommand{\Firststage}{Surgical\xspace}
\newcommand{\secondstage}{structure-aware\xspace}

\newcommand{\weizzOne}{{\ourname}$^{\dagger}$\xspace}
\newcommand{\weizzTwo}{{\ourname}$^{\ddagger}$\xspace}

\newcommand{\rev}[1]{{\color{black} #1}} 
\newcommand{\okrev}[1]{{\color{black} #1}} 

\newlength{\textfloatsepsave}
\newlength{\floatsepsave}
\hypersetup{
  colorlinks   = true, 
  urlcolor     = blue, 
  linkcolor    = black, 
  citecolor   = black 
}

\lstnewenvironment{code}[1][]{
  \noindent
  \minipage{\linewidth}
  \vspace{0.5\baselineskip}
  \lstset{language=C,
    basicstyle=\footnotesize\ttfamily,
    breaklines=true,
    showstringspaces=false
  }}
{
\endminipage
}

\lstnewenvironment{scozzo}[1][]{
  \noindent
  \minipage{\linewidth}
  \vspace{0.5\baselineskip}
  \lstset{language=C,
    basicstyle=\scriptsize\ttfamily,
    breaklines=true,
    xleftmargin=2mm,
    showstringspaces=false
  }}
{
\endminipage
}

\newcommand\daio{} 

\begin{document}

\title[WEIZZ: Automatic Grey-Box Fuzzing for Structured Binary Formats]{WEIZZ: Automatic Grey-Box Fuzzing \\ for Structured Binary Formats}


\newcommand{\marco}[1]{\underline{\sc #1}} 
\newcommand{\edit}[1]{{\color{black} #1}} 
\newcommand{\ppuh}[1]{{\color{red} \sout{#1}}}

\begin{abstract}
Fuzzing technologies have evolved at a fast pace in recent years, revealing bugs in programs with ever increasing depth and speed. Applications working with complex formats are however more difficult to take on, as inputs need to meet certain format-specific characteristics to get through the initial parsing stage and reach deeper behaviors of the program.

Unlike prior proposals based on manually written format specifications, we propose a technique to automatically generate and mutate inputs for unknown chunk-based binary formats. We identify dependencies between input bytes and comparison instructions, and use them to assign tags that characterize the processing logic of the program. Tags become the building block for structure-aware mutations involving chunks and fields of the input. 

Our technique can perform comparably to structure-aware fuzzing proposals that require human assistance. Our prototype implementation \ourname revealed 16 unknown bugs in widely used programs.
\end{abstract}





\author{Andrea Fioraldi} 
    \affiliation{ 
      \institution{Sapienza University of Rome}
      \country{Italy}
    }
    \email{andreafioraldi@gmail.com}

\author{Daniele Cono D'Elia} 
    \affiliation{ 
      \institution{Sapienza University of Rome}
      \country{Italy}
    }
    \email{delia@diag.uniroma1.it}
    
\author{Emilio Coppa} 
    \affiliation{ 
      \institution{Sapienza University of Rome}
      \country{Italy}
    }
    \email{coppa@diag.uniroma1.it}

\maketitle


\section{Introduction}
\label{se:introduction}



Recent years have witnesses a spike of activity in the development of efficient techniques for fuzz testing, also known as fuzzing. In particular, the coverage-based grey-box fuzzing (CGF) approach has proven to be very effective for finding bugs often indicative of weaknesses from a security standpoint. 
The availability of the AFL fuzzing framework~\cite{afl} paved the way to a large body of works proposing not only more efficient implementations, but also new techniques to deal with common fuzzing roadblocks represented by magic numbers and checksum tests in programs. Nonetheless, there are still several popular application scenarios that make hard cases even for the most advanced CGF fuzzers.

CGF fuzzers operate by mutating inputs at the granularity of their bit and byte representations, deeming mutated inputs interesting when they lead execution to explore new program portions. While this approach works very well for compact and unstructured inputs~\cite{superion}, it can lose efficacy for highly structured inputs that must conform to some grammar \okrev{or other type} of specification.

Intuitively, ``undisciplined'' mutations make a fuzzer spend important time in generating many inputs that the initial parsing stages of some program typically rejects, resulting in little-to-none code coverage improvement. Researchers thus have added a user-supplied specification to the picture to produce and prioritize meaningful inputs: \okrev{enhanced CGF} embodiments of this kind are available for both grammar-based~\cite{superion,nautilus} and chunk-based~\cite{aflsmart-tse} formats\edit{, with the latter seeming prevalent among real-world software~\cite{aflsmart-tse}}.

The shortcomings of this approach are readily apparent. Applications typically do not come with format specifications suitable to this end. Asking users to write one is at odds with a driving factor behind the success of CGF fuzzers, that is, they work with a minimum amount of prior knowledge~\cite{redqueen}. Such a request can be expensive, and hardly applies to security contexts where users deal with proprietary or undocumented formats~\cite{grimoire}.
The second limitation is that testing only inputs perfectly adhering to the specification would miss imprecisions in the implementation, while inputs that are to some degree outside it may instead exercise them~\cite{grimoire}. 



\smallskip
{\bf Our approach.} The main feature of our proposal can be summarized as: {\em we attempt to learn how \edit{some chunk-based} input structure may look like based on how \edit{a} program handles its bytes}.

Nuances of this idea are present in~\cite{grimoire} where code coverage guides grammar structure inference, and to some extent in a few general-purpose fuzzing additions. For instance, taint tracking can reveal which input bytes take part in comparisons with magic sequences~\cite{vuzzer,angora}, while input-to-state relationships~\cite{redqueen} can identify also checksum fields using values observed as comparison operands.

We build on the intuition that among comparisons operating on input-derived data, we can deem some as \okrev{tightly coupled} to a single portion of the input structure. We also experimentally observed that the order in which a program exercises these checks can reveal structural details such as the location for the type specifier of a chunk. We tag input bytes with the \okrev{most representative comparison for their processing}, and heuristically infer plausible boundaries for chunks and fields. With this automatic pipeline, we can apply structure-aware mutations for chunk-based grey-box fuzzing~\cite{aflsmart-tse} without the need for a user-supplied format specification.

We start by determining candidate instructions that we can consider relevant with respect to an input byte. Instead of resorting to taint tracking, we flip bits in every input byte, execute the program, and build a dependency vector for operands at comparison instruction sites. We analyze dependencies to identify input-to-state relationships and roadblocks, and to assign tags to input bytes. Tag assignment builds on spatial and temporal properties, as a program typically uses distinct instructions in the form of comparisons to parse distinct items. To break ties we prioritize older instructions as format validation normally happens early in the execution. %
%
Tags drive the inference process of an approximate chunk-based structure of the input, enabling subsequent structure-aware mutations.


\edit{Unlike reverse engineering scenarios} for reconstructing specifications of input formats~\cite{tupni,zeller-dta,autogram}, experimental results suggest that \edit{in this context an} inference process does not have to be fully accurate: fuzzing can deal with some amount of noise and imprecision. Another important consideration is that a specification does not prescribe how its implementation should look like. Developers can share code among functionalities, introducing subtle bugs: we have observed this phenomenon for instance in applications that operate on multiple input formats. Also, they can devise optimized variants of one functionality, specialized on e.g. value ranges of some relevant input field. \okrev{As we assign tags and attempt inference over one input instance at a time, we have a chance to} identify and exercise such patterns when aiming for code coverage improvements.

\smallskip
\edit{{\bf Contributions.} By using information that is within immediate reach of a fuzzer in its usual work, we propose a method to infer an approximate structure for a chunk-based input and then mutate it in a fully automatic manner. Throughout the paper we present:}
\begin{itemize}
\item a dependency identification technique embedded in the deterministic mutation stage of grey-box fuzzing;
\item a tag assignment mechanism that uses such dependencies to overcome fuzzing roadblocks and to back a structure inference scheme for chunk-based formats;
\item an implementation of the approach called \ourname.
\end{itemize}

\noindent
In our experiment\edit{s} \ourname beats or matches a chunk-based CGF proposal that requires a \edit{format} specification, and outperforms \edit{several} general-purpose fuzzers over \okrev{the applications we considered}. We make \ourname available \edit{as open source} at:

\begin{center}
{\tt \url{https://github.com/andreafioraldi/weizz-fuzzer}}
\end{center}



\section{State of the Art}
\label{se:related}

The last years have seen a large body of fuzzing-related works~\cite{fuzzing-book}. Grey-box fuzzers have gradually replaced initial {\em black-box} fuzzing proposals where mutation decisions happen without taking into account their impact on the execution path~\cite{fuzzing-intro}. Coverage-based {\em grey-box} fuzzing (CGF) uses lightweight instrumentation to measure code coverage, which discriminates whether mutated inputs are sufficiently ``interesting'' by looking at control-flow changes.

CGF is very effective in discovering bugs in real software~\cite{aflsmart-tse}, but may struggle in the presence of roadblocks (like magic numbers and checksums)~\cite{redqueen}, or when mutating structured grammar-based~\cite{grimoire} or chunk-based~\cite{aflsmart-tse} input formats. In this section we will describe the workings of the popular CGF engine that \ourname builds upon, and recent proposals that cope with the challenges listed above. We conclude by discussing where \ourname fits in this landscape.


\subsection{Coverage-Based Fuzzing}
\label{ss:related-cgf}
\vspace{-2pt}


American Fuzzy Lop (\afl)~\cite{afl} \okrev{is a very well-known CGF fuzzer and several research works have built on it to improve its effectiveness}. To build new inputs, \afl draws from a queue of initial user-supplied {\em seeds} and previously generated inputs, and runs two mutation stages. Both the {\em deterministic} and the {\em havoc} nondeterministic stage look for coverage improvements and crashes. \afl adds to the queue inputs that improve the coverage of the program, and reports crashing inputs to users as proofs for bugs. 

{\bf Coverage.} AFL adds instrumentation to a binary program to intercept when a branch is hit during the execution. To efficiently track {\em hit counts}, it uses a coverage map of branches indexed by a hash of its source and destination basic block addresses. \afl deems an input {\em interesting} when the execution path reaches a branch that yields a previously unseen hit count. To limit the number of monitored inputs and discard likely similar executions, hit counts undergo a normalization step (power-of-two buckets) for lookup.

{\bf Mutations.} The deterministic stage of \afl sequentially scans the input, applying for each position a set of mutations such as bit or byte flipping, arithmetic increments and decrements, substitution with common constants (e.g., 0, -1, {\tt MAX\_INT}) or values from a user-supplied {\em dictionary}. \afl tests each mutation in isolation by executing the program on the derived input and inspecting the coverage, then it reverts the change and moves to another. 

The havoc stage of \afl applies a nondeterministic sequence (or {\em stack}) of mutations before attempting execution. Mutations come in a random number (1 to 256) and consists in flips, increments, decrements, deletions, etc. at a random position in the input. 

\vspace{-1ex}
\subsection{Roadblocks}
\label{ss:related-roadblocks}
\vspace{-1pt}

{\em Roadblocks} are comparison patterns over the input that are intrinsically hard to overcome through blind mutations: the two most common embodiments are {\em magic numbers}, \okrev{often found for instance in header fields, and {\em checksums}, typically used to verify data integrity}. 
Format-specific dictionaries may help with magic numbers, yet the fuzzer has to figure out where to place such values. Researchers over the years have come up with a few approaches to handle magic numbers and checksums in grey-box fuzzers.

{\bf Sub-instruction Profiling.} While understanding how a large amount of logic can be encoded in a single comparison is not trivial, \okrev{one could} break multi-byte comparisons into single-byte~\cite{lafintel} (or even single-bit~\cite{makingsoftwaredumberer}) checks to better track progress when attempting to match constant values. \lafintel~\cite{lafintel} and {\sc CompareCoverage}~\cite{compcov} are compiler extensions to produce binaries for such a fuzzing. \hongfuzz~\cite{hongg} implements this technique for fuzzing programs when the source is available, while \aflpp~\cite{aflpp} can automatically transform binaries during fuzzing. \steelix~\cite{Li:2017:SPB:3106237.3106295} resorts to static analysis to filter out likely uninteresting comparisons from profiling. While sub-instruction profiling is valuable to overcome tests for magic numbers, it is ineffective however with checksums.

\vspace{1pt}
{\bf Taint Analysis.} Dynamic taint analysis (DTA)~\cite{sok-DTA} tracks when and which input parts affect program instructions. \vuzzer~\cite{vuzzer} uses DTA for checks on magic numbers in binary code, identified as comparison instructions where an operand is input-dependent and the other is constant: \vuzzer places the constant value \okrev{in the input portion that propagates directly to the former operand}. \angora~\cite{angora} brings two improvements: it identifies magic numbers that are not contiguous in the input with a multi-byte form of DTA, and uses gradient descent to mutate tainted input bytes efficiently. \okrev{It however requires compiler transformations to enable such mutations.} 

{\bf Symbolic Execution.} DTA techniques can at best identify checksum tests, but not provide sufficient information to address them. Several proposals (e.g., \cite{stephens2016driller,Yun:2018:QPC:3277203.3277260,conf/sp/WangWGZ10, t-fuzz}) use symbolic execution~\cite{SurveySymbolic} to identify and try to solve complex input constraints involved in generic roadblocks. \taintscope~\cite{conf/sp/WangWGZ10} identifies possible checksums with DTA, patches them away for the sake of fuzzing, and later tries to repair inputs with symbolic execution. It requires the user to provide specific seeds and is subject to false positives~\cite{stephens2016driller,t-fuzz}.


\edit{\tfuzz~\cite{t-fuzz} makes a step further by removing the need for specific seeds and extends the set of disabled checks during fuzzing to any sanity ({\em non-critical}) check that is hard to bypass for a fuzzer but not essential for the computation. For instance, magic numbers can be safely ignored to help fuzzing, while a check on the length of a field should not be patched away. As in \taintscope, crashing inputs are repaired using symbolic execution or manual analysis.}

\driller~\cite{stephens2016driller} instead uses symbolic execution as an alternate strategy to explore inputs, switching to it when fuzzing exhausts a budget obtaining no coverage improvement. Similarly, \qsym~\cite{Yun:2018:QPC:3277203.3277260} builds on concolic execution implemented via dynamic binary instrumentation~\cite{sok-DBI} to trade \okrev{exhaustiveness} for speed.

\smallskip 
{\bf Approximate Analyses.} A recent trend is to explore solutions that approximately extract the information that DTA or symbolic execution can bring, but faster. 

\redqueen~\cite{redqueen} builds on the observation that often input bytes flow directly, or after simple \okrev{encodings} (e.g. swaps for endianness), into instruction operands. This {\bf input-to-state} (I2S) correspondence can be used instead of DTA and symbolic execution to deal with magic bytes, multi-byte compares, and checksums. \edit{ \redqueen approximates DTA by changing input bytes with random values ({\em colorization}) to increase the entropy in the input and then looking for matching patterns between comparisons operands and input parts, which would suggest a dependency.} When both operands of a comparison change, but only for one there is an I2S relationship, \redqueen deems it as a likely checksum test. It then patches the operation, and later attempts to repair the input with educated guesses consisting in mutations of the observed comparison operand values. A topological sort of patches addresses nested checksums.

\edit{\slf~\cite{SLF} attemps a dependency analysis to deal with the generation of valid input seeds when no meaningful test cases are available. It starts from a small random input and flips individual bits in each byte, executing the program to collect operand values involved in comparisons. When consecutive input bytes affect the same set of comparisons across the mutations, they are marked as part of the same field. \slf then uses heuristics on observed values to classify checks according to their relations with the exercised inputs: its focus are identifying offsets, counts, and length of input fields, as they can be difficult to reason about for symbolic executors.}

In the context of concolic fuzzers, \eclipser~\cite{ECLIPSER-ICSE19} relaxes the path constraints over each input byte. It runs a program multiple times by mutating individual input bytes to identify the affected branches. Then it collects constraints resulting from them, considering however only linear and monotone relationships, as other constraint types would likely require a full-fledged SMT solver. \eclipser then selects one of the branches and flips its constraints to generate a new input, mimicking dynamic symbolic execution~\cite{SurveySymbolic}.

\begin{table}[t]
  \centering
  \caption{Comparison with related approaches.\label{tab:related}}
  \vspace{-2mm}
  \adjustbox{max width=\columnwidth}{
  \setlength\tabcolsep{8pt} 
  \begin{tabular}{ c c c c c c c }
    \hline
\thead{\normalsize Fuzzer} & \thead{\normalsize Binary} & \thead{\normalsize Magic bytes} & \thead{\normalsize Checksums} & \thead{\normalsize Chunk-based} & \thead{\normalsize Grammar-based} & \thead{\normalsize \okrev{Automatic}}\\
    \hline

    \aflpp & \cmark & \cmark & \xmark & \xmark & \xmark & \cmark \\
    \angora & \xmark & \cmark & \xmark & \xmark & \xmark & \cmark \\
    \eclipser & \cmark & \cmark & \xmark & \xmark & \xmark & \cmark \\
    \redqueen & \cmark & \cmark & \cmark & \xmark & \xmark & \cmark \\
    \steelix & \cmark & \cmark & \xmark & \xmark & \xmark & \cmark \\

    \nautilus & \xmark & \cmark & \xmark & \xmark & \cmark & \xmark \\
    \superion & \cmark & \cmark & \xmark & \xmark & \cmark & \xmark \\
    \grimoire & \cmark & \cmark & \cmark & \xmark & \cmark & \cmark \\

    \aflsmart & \cmark & \cmark & \xmark & \cmark & \xmark & \xmark \\

    \hline
    \ourname & \cmark & \cmark & \cmark & \cmark & \xmark & \cmark \\
    \hline
  \end{tabular}
  }
\end{table}

\vspace{-1pt}
\vspace{-1ex}
\subsection{Format-Aware Fuzzing}
\label{ss:related-format}
\vspace{-1pt}

Classic CGF techniques lose part of their efficacy when dealing with structured input formats found in files. As mutations happen on bit-level representations of inputs, they can hardly bring the structural changes required to explore new compartments of the data processing logic of an application. Format awareness can however boost CGF: in the literature we can distinguish techniques targeting {\em grammar-based} formats, where inputs comply to a language grammar, and {\em chunk-based} ones, where inputs follow a tree hierarchy with C structure-like data chunks to form individual nodes. 

\vspace{1pt}
{\bf Grammar-Based Fuzzing.} \langfuzz~\cite{Holler:2012:FCF:2362793.2362831} generates valid inputs for a Javascript interpreter using a grammar, combining in a black-box manner sample code fragments and test cases.
 \nautilus~\cite{nautilus} and \superion~\cite{superion} are recent grey-box fuzzer proposals that can test \okrev{language interpreters} without requiring a large corpus of valid inputs or fragments, but only an ANTLR grammar file. 

\grimoire~\cite{grimoire} then removes the grammar specification requirement. Based on \redqueen, it identifies fragments from an initial set of inputs that trigger new coverage, and strips them from parts that would not cause a coverage loss. \grimoire notes information for such gaps, and later attempts to recursively splice-in parts seen in other positions, \okrev{thus mimicking grammar combinations}.


\vspace{1pt}
{\bf Chunk-Based Fuzzing.} 
\spike~\cite{spike} lets users describe the network protocol in use for an application to improve black-box fuzzing. \peach~\cite{peach} generalizes this idea, applying format-aware mutations on an initial set of valid inputs using a user-defined input specification dubbed {\em peach pit}. As they are input-aware, some literature calls such black-box fuzzers {\em smart}~\cite{aflsmart-tse}. However a smart grey-box variant may outperform them, \okrev{as due to the lack of feedback (as in explored code)} they do not keep mutating interesting inputs.

\aflsmart~\cite{aflsmart-tse} validates this speculation by adding smart (or {\em high-order}) mutations to \afl that add, delete, and splice chunks in an input. Using a peach pit, \aflsmart maintains a {\em virtual structure} of the current input, represented conceptually by a tree whose internal nodes are chunks and leaves are attributes. Chunks are characterized by initial and end position in the input, a format-specific chunk type, and a list of children attributes and nested chunks. An attribute is a \okrev{{\em field}} that can be mutated without altering the structure. Chunk addition involves adding as sibling a chunk taken from another input, with both having a parent node of the same type. Chunk deletion trims the corresponding input bytes. Chunk splicing replaces data in a chunk using a chunk of the same type from another input. As virtual structure construction is expensive, \aflsmart defers smart mutations based on the time elapsed since it last found a new path, as trying them over every input would make \aflsmart fall behind classic grey-box fuzzing. 

\subsection{Discussion}
\label{ss:related-discussion}

Tables~\ref{tab:related} depicts where \ourname fits in the state of the art of CGF techniques mentioned in the previous sections. Modern general-purpose CGF fuzzers (for which we choose a representative subset with the first five entries) can handle magic bytes, but only \redqueen proposes \okrev{an effective} solution for generic checksums. In the context of grammar-based CGF fuzzers, \grimoire is currently the only fully automatic (i.e., no format specification needed) solution, and can handle both magic bytes and checksums thanks to the underlying \redqueen infrastructure. With \weizz we bring similar enhancements to the context of chunk-based formats, proposing a scheme that at the same time can handle magic bytes and checksums and eliminates the need for a specification that characterizes \aflsmart.



\section{Methodology}
\label{se:methodology}

\edit{The fuzzing logic of \ourname comprises two stages, depicted in Figure~\ref{fig:weizz-overview}. Both stages pick from a shared queue made of inputs processed by previous iterations of either stage. The surgical stage  identifies dependencies between an input and the comparisons made in the program: it summarizes them by placing {\em tags} on input bytes, and applies deterministic mutations to the sole bytes that turn out to influence operands of comparison instructions. The structure-aware stage extends the nondeterministic working of \afl, leveraging previously assigned tags to infer the location of fields and chunks in an input and mutate them. In the next sections we will detail the inner workings of the two stages.
}

\subsection{\Firststage Stage}
\label{ss:first-stage}

\edit{
The surgical stage replaces the deterministic working of AFL, capturing in the process dependency information for comparison instructions that is within immediate reach of a fuzzer. \ourname summarizes it for the next stage by placing tags on input bytes, and uses it also to identify I2S relationships (Section~\ref{ss:related-roadblocks}) and checksums.

The analysis is context-sensitive, that is, when analyzing a comparison we take into account its calling context~\cite{hcct-spe16}: the {\em site} of the comparison is computed as the exclusive OR of the instruction address with the word used to encode the calling context.

\ourname also maintains a global structure $CI$ to keep track of comparison instructions that the analysis of one or more inputs indicated as possibly involved in checksum tests.

We will use the running example of Figure~\ref{fig:example-surgical} to complement the description of each component of the surgical stage. Before detailing them individually, we provide the reader with an overview of the workflow that characterizes this stage as an input enters it.
}

\subsubsection{Overview}
Given an input $I$ picked from the queue, \ourname{} \edit{attempts} to determine the dependencies between every bit of \edit{$I$} taken individually and the comparison instructions in the program.


Procedure {\sc GetDeps} builds two data structures, both indexed by a hash function $Sites$ for comparison sites. \okrev{A comparison table} $CT$ stores values for operands involved in observed instances of comparison instructions at different sites. For such operands $Deps$ stores which bytes in the input can influence them.  

\ourname then moves to analyzing the recorded information. For each site, it iterates over each instance present in $CT$ looking for I2S dependencies ({\sc DetectI2S}$\,\mapsto\,${\sc R}) and checksum information ({\sc MarkChecksums}$\,\mapsto\,${\sc CI}), and mutates bytes that can alter comparison operands leveraging recorded values ({\sc FuzzOperands}). 

The stage then moves to tag construction, with each input byte initially untagged. It sorts comparison sites according to when they were first encountered and processes them in this order.
Procedure \marco{PlaceTags} assigns a tag to each input byte taking into account I2S relationships $R$, checksum information $CI$, the initially computed dependencies $Deps$, and data associated with comparison sites $CT$. 

Like other fuzzers~\cite{redqueen,conf/sp/WangWGZ10} \ourname forces checksums by patching involved instructions as it detects one, postponing to the end of the surgical stage the input repairing process \marco{FixChecksums}
 required to meet the unaltered checksum condition. To this end we use a technique very similar to the one of \redqueen (Section~\ref{ss:related-roadblocks}).

The output of the surgical stage is an input annotated with discovered tags and repaired to meet checksums, and enters the queue ready for the \secondstage stage to pick it up. 

\edit{The procedures for which we report the name with \underline{underlined} style are described only informally in this paper: the reader can find their full pseudocode in our extended online technical report\footnotemark[1].}

\footnotetext[1]{\url{https://arxiv.org/abs/1911.00621} (contents in \textsection\ref{se:apx}).}




\begin{figure}[t!]
\centering
\includegraphics[width=0.95\linewidth]{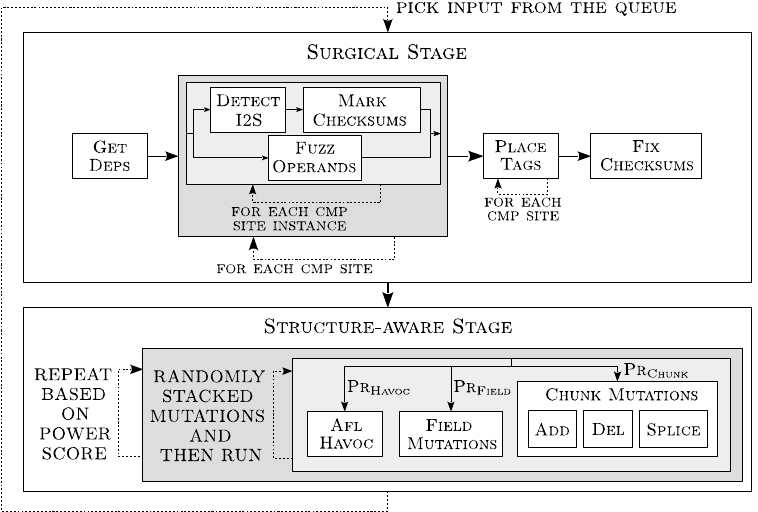}
\vspace{-2.5mm}
\captionof{figure}{Two-stage architecture of {\ourname}.}
\label{fig:weizz-overview}
\end{figure}

\begin{figure*}[t]
\begin{minipage}{.79\columnwidth}
\vspace{-3mm}
\begin{subfigure}{1.0\columnwidth}
\begin{scozzo}
    void parser(char* input, int len) {
       unsigned short id, size, expected, ck = 0;
       int i, offset = 0;
       id = read_ushort(input + offset);
       offset += sizeof(id);
       size = read_ushort(input + offset);
       offset += sizeof(size);
CMP_A: if (id >= 0xAAAA) exit(1);
CMP_B: if (size > len - 3*sizeof(short)) exit(1);
       offset += size;
CMP_C: for (i = 0; i < offset; ++i)
           ck ^= input[i] << (i % 8);
       expected = read_ushort(input + offset);
CMP_D: if (ck != expected) exit(1);

       // process id and data
    }
\end{scozzo}\\\vspace{-4mm}
\caption{\label{fig:example-surgical-a}}
\end{subfigure}
\end{minipage}
\begin{minipage}{1.3\columnwidth}
\begin{minipage}{1.0\columnwidth}
\begin{subfigure}{1.0\columnwidth}
\begin{minipage}{0.4\columnwidth}
\vspace{-3mm}
\caption{\label{fig:example-surgical-b}}
\end{minipage}
\begin{minipage}{0.6\columnwidth}\hspace*{-20mm}
\begin{footnotesize}
\begin{tabular}{ll}
Input format: & {\tt [id] [size] [data] [checksum]} \\ 
where: & {\tt id}, {\tt size}, and {\tt checksum} are 2-byte long \\
       & {\tt data} is {\tt size}-byte long \\
\end{tabular}\\
\end{footnotesize}
\end{minipage}
\medskip
\end{subfigure}
\end{minipage}
\begin{minipage}{0.5\columnwidth}
\begin{subfigure}{1.0\columnwidth}
\centering
\begin{footnotesize}
Bytes of seed:\\
{\tt [0E 00][02 00][41 41][36 0C]}
\\\vspace{1.5mm}
Collected comparison instances:\\
\begin{tabular}{|c|c|c|c|}\hline
 {\tt CMP}$_{\tt A}$ & {\tt CMP}$_{\tt B}$ & {\tt CMP}$_{\tt C}$ & {\tt CMP}$_{\tt D}$ \\ \hline\hline
 {\tt  E, AAAA} & {\tt 2, 2} & {\tt 0, 6} & {\tt C36, C36} \\ \hline
              &            & {\tt 1, 6} & \\ \hline
              &            & {\tt 2, 6} & \\ \hline
              &            & {\tt 3, 6} & \\ \hline
              &            & {\tt 4, 6} & \\ \hline
              &            & {\tt 5, 6} & \\ \hline
              &            & {\tt 6, 6} & \\ \hline
\end{tabular}\\\vspace{-1.0mm}
\end{footnotesize}
\caption{\label{fig:example-surgical-c}}
\end{subfigure}
\end{minipage}
\begin{minipage}{.5\columnwidth}
\begin{subfigure}{1.0\columnwidth}
\centering
\begin{footnotesize}
Bytes of seed after flipping first bit:\\
{\tt [\textbf{8}E 00][02 00][41 41][36 0C]}
\\\vspace{1.5mm}
Collected comparison instances:\\
\begin{tabular}{|c|c|c|c|c|}\hline
 {\tt CMP}$_{\tt A}$ & {\tt CMP}$_{\tt B}$ & {\tt CMP}$_{\tt C}$ & {\tt CMP}$_{\tt D}$ \\ \hline\hline
 {\tt  \textbf{8E}, AAAA} & {\tt 2, 2}& {\tt 0, 6} & {\tt \textbf{CB6}, C36} \\ \hline
              &            & {\tt 1, 6} & \\ \hline
              &            & {\tt 2, 6} & \\ \hline
              &            & {\tt 3, 6} & \\ \hline
              &            & {\tt 4, 6} & \\ \hline
              &            & {\tt 5, 6} & \\ \hline
              &            & {\tt 6, 6} & \\ \hline
\end{tabular}\\\vspace{-1.0mm}
\end{footnotesize}
\caption{\label{fig:example-surgical-d}}
\end{subfigure}
\end{minipage}
\end{minipage}
\vspace{-3mm}
\caption{\edit{Example for surgical stage: (a) code of function {\tt parser}; (b) input format; comparison instances collected by \ourname when running {\tt parser} on (c) the initial seed and (d) on the seed after flipping its first bit  {\em (affected operands are marked in bold)}.\label{fig:example-surgical}}}
\end{figure*}

\subsubsection{Dependency Identification}
\label{ss:get-deps}


\begin{figure}[t]
\removelatexerror
\begin{algorithm}[H]
\setlength{\textfloatsep}{0mm}
\footnotesize
  \DontPrintSemicolon
  \SetAlgoNoEnd
  \SetAlgoNoLine
  \SetNlSkip{-0.45em}
  \SetKwFunction{FMain}{\sc GetDeps}
  \SetKwProg{Fn}{function}{:}{}
  {\scriptsize $CT$: $Sites\times J \times \{\textit{op}_1\textit{, op}_2\}$ $\to$ $V$   [cmp site instance \& operand$\,\to\,$value]}\;
  {\scriptsize $Deps$: $Sites\times J \times \{\textit{op}_1\textit{, op}_2\}$ $\to$ $A$  [cmp site inst. \& operand$\,\to$ array of |input| bools]}\;
  
  \Fn{\FMain{I}}{
  \Indmm
    \nextnr CT $\gets$ {\sc RunInstr}(I)\;
    \nextnr \ForEach{b$\,\in\,$\{0\,...\,len(I)-1\}}{\Indmm
      \nextnr Deps(s, j, op)[b] $\gets$ false $\forall\,$(s, j, op) $\in$ dom(CT)\;
      \nextnr \ForEach{k $\in$ \{0\,...\,7\}}{\Indmm
          \nextnr CT$^{\prime}$ $\gets$ {\sc RunInstr}({\sc BitFlip}(I, b, k))\;
          \nextnr \ForEach{(s, j, op) $\in$ dom(CT)}{\Indmm
              \nextnr \lIf{{\sc Hits}(CT, s) $\neq$ {\sc Hits}(CT$^{\prime}$, s)}{{\bf continue}}
          \nextnr Deps(s, j, op)[b] $\gets$ Deps(s, j, op)[b] $\vee$ CT(s, j, op)$\,\neq\,$CT$^{\prime}$(s, j, op)\; 
      }
      }
    }
    \nextnr \Return{CT, Deps}\;
  }
  \vspace{1.2mm}
  \caption{Dependency identification step.\label{alg:weizz-deps}}
\end{algorithm}
\setlength{\textfloatsep}{\textfloatsepsave}
\vspace{-2.5mm}
\end{figure}

A crucial feature that makes a fuzzer effective is the ability to understand how mutations over the input affect the execution of a program. In this paper we go down the avenue of fast analyses to extract approximate dependency information. We propose a technique that captures which input bytes individually affect comparison instructions: it can capture I2S facts like the technique of \redqueen, but also non-I2S relationships that turn out to be equally important to assign tags to input bytes.

Algorithm~\ref{alg:weizz-deps} describes the {\sc GetDeps} procedure used to this end. Initially we run the program over the current input instrumenting all comparison instructions. The output is a comparison table $CT$ that records the operands for the most recent $|J|$ \okrev{observations ({\em instances}) of} a comparison site. For each comparison site $CT$ keeps a timestamp for when {\sc RunInstr} first observed it, and how many times it encountered such site in the execution ({\sc Hits} at line 7). 



\ourname then attempts to determine which input bytes contribute to operands at comparison sites, either directly or through derived values. By altering bytes individually, \ourname looks for sites that see their operands changed after a mutation. The algorithm iterates over the bits of each byte, flipping them one at a time and obtaining a $CT'$ for an instrumented execution with the new input (line 5). We mark the byte as a dependency for an operand of a comparison site instance (line 8) if its $b$-th byte has changed in $CT'$ w.r.t. $CT$.

Bit flips may however cause execution divergences, as some comparisons likely steer the program towards different paths (and when such an input is {\em interesting} we add it to the queue, see Section~\ref{ss:related-cgf}). Before updating dependency information, we check whether a comparison site $s$ witnessed a different number of instances in the two executions (line 7). This is a proxy to determine whether execution did not diverge locally at the comparison site. We prefer a local policy over enforcing that all sites in $CT$ and $CT'$ see the same number of instances. In fact for a mutated input some code portions can see their comparison operands change (suggesting a dependency) without incurring control-flow divergences, while elsewhere the two executions \okrev{may} differ too much to look for correlations.

\boxedspit{
We discuss a simple parsing code (Figure~\ref{fig:example-surgical-a}) for a custom input format characterized by three fields {\tt id}, {\tt size}, and {\tt checksum}, each expressed using 2 bytes, and a field {\tt data} of variable length encoded by the {\tt size} field (Figure~\ref{fig:example-surgical-b}).

Our parser takes as argument a pointer {\tt input} to the incoming bytes and their number {\tt len}. It assumes the input buffer will contain at least 6 bytes, which is the minimum size for a valid input holding empty data. The code operates as follows:
\begin{enumerate}[label=(\alph*),leftmargin=23pt]
\item it reads the {\tt id} and {\tt size} fields from the buffer;
\item it checks the validity of field {\tt id} (label {\tt CMP\_A});
\item it checks whether the buffer contains at least 6+{\tt size} bytes to host {\tt data} and the other fields (label {\tt CMP\_B});
\item it computes a checksum value iterating (label {\tt CMP\_C}) over the bytes associated with {\tt id}, {\tt size}, and {\tt data};
\item it reads the {\tt expected} checksum from the buffer and validates the one presently computed (label {\tt CMP\_D}).
\end{enumerate}


Figure~\ref{fig:example-surgical-c} shows the comparison instances collected by \ourname when running the function {\tt parser} over the initial seed. For brevity we omit timestamps and hit counts and we assume that the calling context is not relevant, so we can use comparison labels to identify sites. Multiple instances appear for site {\tt CMP\_C} as it is executed multiple times within a loop.


Figure~\ref{fig:example-surgical-d} shows the comparison instances collected after flipping one bit in the first byte of the seed, with changes highlighted in bold. Through these changes \ourname detects that the first operand of both {\tt CMP\_A} and {\tt CMP\_D} was affected, revealing dependencies between these two comparison sites and the first input byte. Subsequent flips will lead to inputs that reveal further dependencies:
\begin{itemize}[leftmargin=17pt]
\item the 1st operand of {\tt CMP\_A} depends on the bytes of {\tt id};
\item the 1st oper. of {\tt CMP\_B} depends on  the bytes of {\tt size};
\item the 2nd oper. of {\tt CMP\_C} depends on the 1st byte of {\tt size}\footnotemark[2];
\item the 1st oper. of {\tt CMP\_D} depends on the bytes of both {\tt id} and {\tt data}, as well as on the first byte of {\tt size}, while the 2nd oper. is affected by the {\tt checksum} field bytes;
\end{itemize}
}

\footnotetext[2]{The dependency is exposed only when flipping the second least significant bit of {\tt size}, turning it into the value zero, as flipping other bits makes the size invalid and aborts the execution prematurely at {\tt CMP\_B}.}

\subsubsection{Analysis of Comparison Site Instances}
\label{ss:checksumsAndFuzz}
After constructing the dependencies, \ourname starts processing the data recorded for up to $|J|$ instances of each comparison site. The first analysis is the~\pagebreak[4] detection of I2S correspondences with {\sc DetectI2S}. \okrev{\redqueen explores this concept to deal with roadblocks, based on the intuition that parts of the input can flow directly into the program state in memory or registers used for comparisons (Section~\ref{ss:related-roadblocks}). For instance, magic bytes found in headers are likely to take part in comparison instructions as-is or after some simple encoding transformation~\cite{redqueen}.} We apply {\sc DetectI2S} to every operand in a comparison site instance and populate the $R$ data structure with newly discovered I2S facts.

\vspace{2pt}
{\sc MarkChecksums} involves the detection of checksumming sequences. \okrev{Similarly to \redqueen, we mark a comparison instruction as} likely involved in a checksum if the following conditions hold:  {\bf (1)} one operand is I2S and its size is at least 2 bytes; {\bf (2)} the other operand is not I2S and \okrev{{\sc GetDeps} revealed} dependencies on some input bytes; and {\bf (3)} $\bigwedge_b$(Deps(s, j, op$_1$)[b],  Deps(s, j, op$_2$)[b]) = false, that is, \okrev{the sets of their byte dependencies} are disjoint.

The intuition is that code compares the I2S operand (which we speculate to be the expected value) against a value derived from input bytes, and those bytes do not affect the I2S operand or we would have a circular dependency. We choose a 2-byte minimum size to reduce false positives.
%
Like prior works \okrev{we patch candidate checksum tests} to be always met, and defer to a later stage both the identification of false positives and the input repairing required to meet the condition(s) from the original unpatched program.

\vspace{2pt}
Finally, {\sc FuzzOperands} replaces the deterministic mutations of AFL with surgical byte substitutions driven by data seen at comparison sites. For each input byte, we determine every CT entry (i.e., each operand of a comparison site instance) it influences. Then we replace the byte and, depending on the operand size, its surrounding ones using the value recorded for the other operand. The replacement can use such value as-is, rotate it for endianness, increment/decrement it by one, perform zero/sign extension, or apply ASCII encoding/decoding of digits. Each substitution yields an input added to the queue \okrev{if its execution reveals} a coverage improvement.

\boxedspit{When analyzing the function {\tt parser}, \ourname is able to detect that the first operands of {\tt CMP\_A} and {\tt CMP\_B} and the second operand of {\tt CMP\_D} are I2S. On the other hand, the second operand of {\tt CMP\_C} and the first operand of {\tt CMP\_D} are not I2S, although they depend on some of the input bytes.

\ourname marks {\tt CMP\_D} as a comparison instance likely involved into a checksum since the three required conditions hold: the second operand is I2S and has size 2, the first operand is not I2S but depends on several input bytes, and the two operands show dependencies on disjoint sets of bytes.

{\sc FuzzOperands} attempts surgical substitutions based on observed operand values: for instance, it can place as-is value {\tt 0xAAAA} recorded at the comparison site {\tt CMP\_A} in the bytes for the field {\tt id} and trigger the enclosed {\tt exit(1)} statement. 
}

\subsubsection{Tag Placement} 
\label{ss:tag-placement}

\ourname succinctly describes dependency information between input bytes and performed comparisons by annotating such bytes with tags essential for the subsequent structural information inference.
%
%
For the $b$-th input byte $Tags[b]$ keeps:
\begin{itemize}[topsep=1pt]
\item {\em id}: (the address of) the comparison instruction chosen as most significant among those $b$ influences;
\item {\em ts}: the timestamp of the {\em id} instruction when first met;
\item {\em parent}: the comparison instruction that led \ourname to the most recent tag assignment prior to $Tags[b]$;
\item {\em dependsOn}: \okrev{when $b$ contains a checksum value, the comparison instruction for the innermost nested checksum that verifies the integrity of $b$, if any;}
\item {\em flags}: stores the operand affected by $b$, if it is I2S, or if it is part of a checksum field;
\item {\em numDeps}: the number of input bytes that the operand in {\em flags} depends on.
\end{itemize}

\noindent
The tag assignment process relies on spatial and temporal locality in how comparison instructions get executed. \ourname tries to infer structural properties of the input based on the intuition that a program typically uses distinct instructions to process structurally distinct items. We can thus tag input bytes as related \rev{when they reach one same comparison directly or through derived data}. When more candidates are found, we use temporal information to prioritize instructions with a lower timestamp, as input format validation normally takes place in parsing code from the early stages of the execution. As we explain next, we extend this scheme \okrev{to account for} checksums and to reassign tags when heuristics spot a likely better candidate than the current choice. \edit{Temporal information can serve also as proxy for hierarchical relationships with parent tags.}


Algorithm \marco{PlaceTags} iterates over comparison sites sorted by when first met in the execution, and attempts an inference for each input byte. \okrev{We apply it to individual $op$ instruction operands seen at a site $s$. If for the current byte $b$ we find no dependency among all recorded instances $Deps(s, j, op)$, the cycle advances to the next byte, otherwise we compute a $numDeps$ candidate, i.e., the number $n$ of input bytes that affect the instruction, computed as $n \gets \sum_k  (1~\textsf{if}\,\bigvee_j Deps(s, j, op)[k]~\textsf{else}~0)$ where $k$ indexes the input length. If the byte is untagged we tag it with the current instruction, otherwise we consider a reassignment. If the current tag $Tags[b]$ does not come from a checksum test and $n$ is smaller than $Tags[b].numDeps$, we reassign the tag as fewer dependencies suggest the instruction may be more representative for the byte}.


When we find a comparison treating the byte as from a checksum value, we always assign it as its tag. To populate the $dependsOn$ field we use a topological sort of the dependencies $Deps$ over each input byte, that is, we know when a byte part of a checksum value represents also a byte that some outer checksum(s) verify for integrity. We later repair inputs starting from the innermost checksum, with $dependsOn$ pointing to the next comparison to process. 

\boxedspit{
Let us consider the seed of Figure~\ref{fig:example-surgical-c}. The analysis of dependencies leads to the following tag assignments: 
\begin{itemize}[leftmargin=17pt]
    \item bytes from field {\tt id} affect sites {\tt CMP\_A} and {\tt CMP\_D}: {\tt CMP\_A} is chosen as tag as it is met first in the execution;
    \item bytes from {\tt size} affect {\tt CMP\_B}, {\tt CMP\_C}, and {\tt CMP\_D}: {\tt CMP\_B} is chosen as tag as it temporally precedes the other sites;
    \item bytes from {\tt data} and {\tt checksum} affect only {\tt CMP\_D}, which becomes the {\em id} for their respective tags: however the tags will differ in the {\em flags} field as those for {\tt checksum} bytes are marked as involved in a checksum field.
\end{itemize}
}



\subsubsection{Checksum Validation \& Input Repair} 
\label{ss:patch-and-fix-checksum}

The last surgical step involves checksum validation and input repair for checksums patched in the program \okrev{during previous iterations}. As the technique is conceptually similar to the one of \redqueen, we discuss it briefly.

Algorithm \marco{FixChecksums} uses topologically sorted tags that were previously marked as involved in checksum fields. For each tag, it first extracts the checksum computed by the program (i.e., the input-derived operand value of the comparison) and where in the input the contents of the other operand (which is necessarily I2S, see Section~\ref{ss:checksumsAndFuzz}) are stored, then it replaces such bytes with the expected value. It then disables the involved patch and runs the program on the repaired input: if it handled the checksum correctly, we will observe the same execution path as before, otherwise the checksum was a false positive and we have to bail.

\okrev{At the end of the process \ourname re-applies patches that were not a false positive: this will benefit both the second stage and future surgical iterations over other inputs. It then also applies patches for checksums newly discovered by {\sc MarkChecksums} in the current surgical stage, so when it will analyze again the input (or similar ones from {\sc FuzzOperands}) it will be able to overcome them.}


\begin{figure}[t]
\centering
\includegraphics[width=0.85\linewidth]{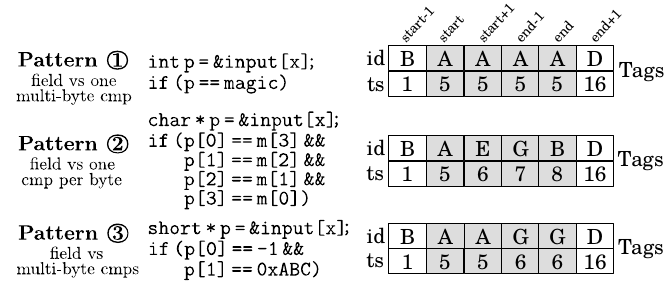}
\vspace{-3mm}
\captionof{figure}{Field patterns identified by \ourname.}
\vspace{-2mm}
\label{fig:field}
\end{figure}

\subsection{Structure-Aware Stage}
\label{se:tag-mutations}

The second stage of \ourname generates new inputs via nondeterministic mutations that can build on tags assigned in the surgical stage. Like in the \afl realm, the number of derived inputs depends on an energy score that the power scheduler of \afl assigns to the original input~\cite{aflsmart-tse}. Each input is the result of a variable number (1 to 256) of stacked mutations, with \ourname picking nondeterministically at each step a {\em havoc}, a {\em field}, or a {\em chunk} mutation scheme.
\okrev{The choice of keeping the havoc mutations of \afl is shared with previous chunk-oriented works~\cite{aflsmart-tse}, as combining them with structure-aware mutations improves the scalability of the approach.} 


\okrev{Field mutations build on tags assigned in the surgical stage to selectively alter bytes that together likely represent a single piece of information. Chunk mutations target instead higher-level, larger structural transformations driven by tags.}

As our field and chunk inference strategy is \okrev{not} sound, especially compared to manually written specifications, we give \ourname leeway in the identification process. \ourname never reconstructs the input structure in full, but only identifies individual fields or chunks in a nondeterministic manner when performing a mutation. The downside of this choice is that \ourname may repeat work or make conflicting choices among mutations, as it does not keep track of past choices. On the bright side, \ourname does not commit to possibly erroneous choices when applying one of its heuristics: this limits their impact to the subsequent mutations, and grants more leeway to the overall fuzzing process in exploring alternate scenarios.

We will describe next how we use input tags to identify fields and chunks and the mutations we apply. At the end of the process, we execute the program over the input generated from the stacked mutations, looking for crashes and coverage improvements. We promptly repair inputs that cause a new crash, while the others undergo repair when they eventually enter the surgical stage.


\removelatexerror
\begin{figure}[t]
\vspace{1mm}
\setlength{\textfloatsep}{0mm}
\begin{algorithm}[H]
\footnotesize
  \DontPrintSemicolon
  \SetAlgoNoEnd
  \SetAlgoNoLine
  \SetNlSkip{-0.45em}
  \SetKwFunction{FMain}{\sc FieldMutation}
  \SetKwProg{Fn}{function}{:}{}
  \Fn{\FMain{Tags, I}}{
  \Indmm
     \nextnr b $\gets$ pick u.a.r. from \{0$\,$...$\,$len(I)\}\;
    \nextnr \For{i $\in$ \{b\,...\,len(I)-1\}}{\Indmm
      \nextnr \lIf{Tags[i].id==0}{{\bf continue}}
      \nextnr start $\gets$ {\sc GoLeftWhileSameTag}(Tags, i)\;
       \nextnr {\bf with} {\em probab. Pr$_{I2S}$} {\bf or} \If{{\sc !I2S}(Tags[start])}{\Indmm
            \nextnr end $\gets$ {\sc FindFieldEnd}(Tags, start, 0)\;
        \nextnr I $\gets$ {\sc Mutate}(I, start, end); \textbf{break}\;
      }
    }
   \nextnr \Return{I}\;
  }
  \vspace{2pt}
  \caption{Field identification and mutation.\label{alg:weizz-field}}
\end{algorithm}
\setlength{\textfloatsep}{\textfloatsepsave}
\vspace{1.6mm}
\end{figure}

\subsubsection{Fields}
\label{ss:fields-mutations}
Field identification is a heuristic process designed \okrev{around} prominent patterns that \ourname should identify. The first one is straightforward and sees a program checking a field using a single comparison instruction. We believe it to be the most common in real-world software. The second one instead sees a program comparing every byte in a field using a different instruction, as in the following fragment from the {\tt lodepng} library:

\vspace{-6pt}
\begin{scozzo}
unsigned char lodepng_chunk_type_equals(const unsigned char* chunk, const char* type) {
  if (strlen(type) != 4) return 0;
  return (chunk[4]==type[0] && chunk[5]==type[1] && chunk[6]==type[2] && chunk[7]==type[3]); }
\end{scozzo}
\vspace{-1pt}

\noindent which checks each byte in the input string {\tt chunk} using a different comparison statement. Such code for instance often appears in program to account for differences in endianness. The two patterns may be combined to form a third one, as in the bottom part of Figure~\ref{fig:field}, that we \okrev{consider in our technique as well}.

\okrev{Let us present how \ourname captures the patterns instantiated in Figure~\ref{fig:field}. For the first pattern,} since a single instruction checks all the bytes in the field, we expect to see the corresponding input bytes marked with the same tag. For the second pattern, we expect instead to have consecutive bytes marked with different tags but consecutive \okrev{associated} timestamps, as no other comparison instruction intervenes. In the figure we can see a field made of bytes with tag ids \{A, E, G, B\} having respective timestamps \{5, 6, 7, 8\}. The third pattern implies having (two or more) subsequences made internally of the same tag, with the tag changing across them but with a timestamp difference of one unit only.

Procedure {\sc FieldMutation} (Algorithm~\ref{alg:weizz-field}) looks nondeterministically for a field by choosing a random position $b$ in the input. If there is no tag for the current byte, the cursor advances until it reaches a tagged byte (line 3). On the contrary, if the byte is tagged \ourname checks whether it is the first byte in the candidate field or there are any preceding bytes with the same tag, retracting to the leftmost in the latter case (line 4). With the start of the field now found, \ourname evaluates whether to mutate it: if the initial byte is I2S, the mutation happens only with a probability as such a byte could represent a magic number. Altering a magic number would lead to program paths handling invalid inputs, which in general are less appealing for a fuzzer. The extent of the mutation is decided by the helper procedure \marco{FindFieldEnd}, which looks for sequences of tagged bytes that meet one of the three patterns discussed above. 
{\sc FieldMutation} then alters the field by choosing one among \okrev{the twelve \afl length-preserving} havoc transformations over its bytes.




\begin{figure}[t]
\adjustbox{width=\linewidth}{
\begin{subfigure}{.34\columnwidth}
\centering
\begin{scozzo}
struct { // lines
 int type; // 1-3
 int x, y; // 2-3
 int cksm; // 4-5
}
\end{scozzo}\vspace{-2.5mm}
\caption{\label{fig:chunk-a}}
\end{subfigure}%
\begin{subfigure}{.32\columnwidth}
\centering
\vspace{-.5mm}
\begin{scozzo}
struct {
 int type; // 1-3
 int cksm; // 4-5
 int x, y; // 8-9
}
\end{scozzo}\vspace{-2.25mm}
\caption{\label{fig:chunk-b}}
\end{subfigure}
\begin{subfigure}{.41\columnwidth}
\centering
\vspace{-.5mm}
\begin{scozzo}
struct { 
 int type; // 1-3
 int x, y; // 2-3
 int cksm; // 4-5
 char data[64]; // 7-9
}
\end{scozzo}\vspace{-5.2mm}
\caption{\label{fig:chunk-c}}
\end{subfigure}
}\\[.4ex]
\begin{subfigure}{\linewidth}
\begin{minipage}{.05\textwidth}
    \centering
    \footnotesize{\bf (d)}
\end{minipage}%
\begin{minipage}{0.95\textwidth}
    \centering
    \includegraphics[width=0.98\linewidth]{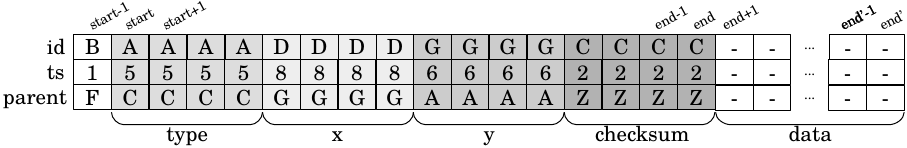}
\end{minipage}
\end{subfigure}
\vspace{-4mm}
\caption{\okrev{Examples of chunks that \ourname can look for.}\label{fig:chunk}}
\vspace{-2mm}
\label{fig:test}
\end{figure}

\subsubsection{Chunks}
\label{ss:chunk-mutations}
\ourname heuristically locates plausible boundaries for chunks in the input sequence, then it applies higher-order mutations to candidate chunks. 
As running example we discuss variants of a structure representative of \okrev{many} chunks found in binary formats.

The first variant (Figure~\ref{fig:chunk-a}) has four fields: a {\tt type} assigned with \okrev{some well-known constant}, two data fields {\tt x} and {\tt y}, and a {\tt cksm} value for integrity checking. Let us consider a program that first computes the checksum of the structure and compares it against the namesake field, then it verifies the {\tt type} and {\tt x} fields in this order, \okrev{executes possibly unrelated comparisons (i.e., they do not alter tags for the structure bytes)}, and later on makes a comparison depending on {\tt y}. For the sake of simplicity we assume that field processing matches the first pattern seen in the previous section (i.e., one comparison per field). The output of {\sc PlaceTags} will be coherent with the graphical representation of Figure~\ref{fig:chunk}d.

We now describe how our technique is able to identify the boundaries of this chunk using tags and their timestamps. Similarly as with fields, we pick a random position $b$ within the input and analyze the preceding bytes, retracting the cursor as long as the tag stays the same. For any $b$ initially falling within the 4 bytes of {\tt type}, \ourname finds the same {\em start} index for the chunk. To find the end position it resorts to {\sc FindChunkEnd} (Algorithm~\ref{alg:weizz-chunk-end}).

The procedure initially recognizes the 4 bytes of {\tt type} (line 1), then recursively looks for adjacent fields when their timestamps are higher than the one for the current field (lines 2-3). This recursive step reveals fields {\tt x} and {\tt y}. The {\tt cksm} field has a lower timestamp value (as the program checked it before analyzing {\tt type}), but lines 4-5 can include it in the chunk as they inspect parent data. The parent is the comparison from the tag assignment prior to the current one, and the first activation of {\sc FindChunkEnd} will see that the tag of the first byte of {\tt cksm} matches the parent for the tag for {\tt type}.

To explain lines 6-9 in Algorithm~\ref{alg:weizz-chunk-end}, let us consider variants of the structure that exercise them. In Figure~\ref{fig:chunk-b} {\tt cksm} comes before {\tt type}, and in this case the algorithm would skip over lines 2-3 without incrementing $end$, thus missing {\tt x} and {\tt y}. Lines 8-9 can add to the chunk bytes from adjacent fields as long as they have increasing timestamps with respect to the one from the tag for the $k$-th byte (the first byte in {\tt type} in this case). In Figure~\ref{fig:chunk-c} we added an array {\tt data} of 64 bytes to the structure: it may happen that \ourname leaves binary blobs untagged \okrev{if the program does not make comparisons over them or some derived values.} Line 7 can add such sequences to a chunk, extending $end$ to the new $end'$ value depicted in Figure~\ref{fig:chunk}d. 

With {\sc FindChunkEnd} we propose an inference scheme inspired by the layout of popular formats. The first field in a chunk is typically also the one that the program analyzes first, and we leverage this fact to speculate where a chunk may start. If the program verifies a checksum before accessing even that field, we link it to the chunk using parent information, otherwise ``later'' checksums represent a normal data field. Lines 7-9 enlarge chunks to account for different layouts and tag orderings, but only with a probability to avoid excessive extension. \okrev{The algorithm can also capture partially tagged blobs through the recursive steps it can take.} 

\okrev{Observe however that in some formats there might be dominant chunk types, and choosing the initial position randomly would reveal a chunk of non-dominant type with a smaller probability.} We devise an alternate heuristic that handles this scenario better: it randomly picks one of the comparison instructions \okrev{appearing} in tags, \okrev{assembles} non-overlapping chunks with {\sc FindChunkEnd} starting at a byte tagged by such an instruction, and picks one among the built chunks. As \ourname is unaware of the format characteristics, we choose between the two heuristics with a probability.


\begin{figure}[t]
\removelatexerror
\setlength{\textfloatsep}{0mm}
\begin{algorithm}[H]
\footnotesize
  \DontPrintSemicolon
  \SetAlgoNoEnd
  \SetAlgoNoLine
  \SetNlSkip{-0.45em}
  \SetKwFunction{FMain}{\sc FindChunkEnd}
  \SetKwIF{With}{ElseWith}{}{with}{do}{else with}{}{endwith}
  \SetKwProg{Fn}{function}{:}{}
  \Fn{\FMain{Tags, k}}{
  \Indmm
    \nextnr end $\gets$ {\sc GoRightWhileSameTag}(Tags, k)\;
    \nextnr \While{Tags[end+1].ts >= Tags[k].ts}{\Indmm
      \nextnr end $\gets$ {\sc FindChunkEnd}(Tags, end+1)\;
    }
    \nextnr \While{Tags[end+1].id == Tags[k].parent}{\nextnr end $\gets$ end +1}
    \nextnr \With{probability Pr$_{\okrev{extend}}$}{\Indmm
      \nextnr \lWhile{Tags[end+1].id == 0}{end $\gets$ end +1}
      \nextnr \While{Tags[end+1].ts >= Tags[k].ts}{\Indmm
        \nextnr end $\gets$ {\sc FindChunkEnd}(Tags, end+1)\;
      }
    }
    \nextnr \Return{end}\;
  }
  \vspace{1pt}
  \caption{Boundary identification for chunks.\label{alg:weizz-chunk-end}}
\end{algorithm}
\setlength{\textfloatsep}{\textfloatsepsave}
\vspace{-2.5mm}
\end{figure}

For the current chunk selection, \ourname attempts one of the following higher-order mutations (Section~\ref{ss:related-format}):
\begin{itemize}[parsep=1pt,topsep=2pt]
  \item {\bf Addition.} It adds a chunk from another input to the chunk that encloses the current one. \ourname picks a tagged input I$^{\prime}$ from the queue and looks for chunks in it starting with bytes tagged with the same parent field of the leading bytes in the current chunk. It picks one randomly and extends the current input by adding its associated bytes before or after the current chunk. The parent tag acts as proxy for the nesting information available instead to \aflsmart.
    \item {\bf Deletion}. It removes the input bytes for the chunk.
  \item {\bf Splicing.} It picks a similar chunk from another input to replace the current one. It scans that input looking for chunks starting with the same tag (\aflsmart uses type information) of the current, randomly picks one, and substitutes the bytes.
\end{itemize}
\vspace{-2pt}



\edit{\subsubsection{Discussion}
We can now elaborate on why, in order to back the field and chunk mutations just described, we cannot rely on the dependency identification techniques of \redqueen or \slf.

Let us assume \redqueen colors an input with a number of {\tt A} bytes and initially logs a {\tt cmp A, B} operation. \redqueen would attempt to replace each occurrence of {\tt A} with {\tt B} and validate it by looking for coverage improvements when running the program over the new input, this time with logging disabled. This strategy works well if the goal are only I2S replacements, but for field identification there are two problems: (i) with multiple {\tt B} bytes in the input, we cannot determine which one makes a field for the comparison, and (ii) if {\tt B} is already in another input in the queue, no coverage improvement comes from the replacement, and \redqueen misses a direct dependency for the current input. Our structural mutations require tracking comparisons all along for I2S and non-I2S facts.

\slf can identify dependencies similarly to our {\sc GetDeps}, but recognizes only specialized input portions---that we can consider fields---based on how the program treats them. \slf supports three categories of program checks  and can mutate portions e.g. by replicating those involved in a count check (\textsection\ref{ss:related-roadblocks}). While \ourname may do that through chunk addition, it also mutates fields---and whole chunks---that are not treated by \slf.}


\vspace{-1pt}
\section{Implementation}
\label{se:implementation}




We implemented our ideas on top of \afl 2.52b and \qemu 3.1.0 for x86-64 Linux targets. For branch coverage and comparison tables we use a shadow call stack to compute context-sensitive information~\cite{hcct-spe16}, which may let a fuzzer explore programs more pervasively~\cite{angora}. We index the coverage map using the source and destination basic block addresses and a hash of the call stack. 


Natural candidates for populating comparison tables are {\tt cmp} and {\tt sub} instructions. We store up to $|J|$=256 entries per site.
Like \redqueen we also record {\tt call} instructions that may involve comparator functions: we check whether the locations for the first two arguments contain valid pointers, and dump 32 bytes for each operand. Treating such calls as operands (think of functions \okrev{that behave} like {\tt memcmp}) may improve the coverage, especially when the fuzzer is configured not to instrument external libraries.



An important optimization involves {\em deferring} surgical fuzzing with a timeout-based mechanism, letting inputs jump to the second stage with a decreasing probability if an interesting input was discovered in the current window (sized as 50 seconds).

Note that untagged inputs can only undergo havoc mutations in the second stage: we thus introduce {\em derived} tags, which are an educated guess on the actual tag based on tags seen in a similar input. \edit{Derived tags speed up our fuzzer, and actual tags replace them when the input eventually enters the surgical stage.

Two scenarios can give rise to derived tags. One case happens when starting from an input I the surgical mutations of {\sc FuzzOperands}(I) produce one or more inputs I' that improve coverage and are added to the queue. Once PlaceTags has tagged I, we copy such tags to I' (structurally analogous to I, as {\sc FuzzOperands} operates with ``local'' mutations) and mark them as derived.

Similarly, a tagged input I1 can undergo high-order mutations that borrow bytes from a tagged input I2, namely addition or splicing. In this case the added/spliced bytes of the mutated I1' coming from I2 get the same tags seen in I2, while the remaining bytes keep the tags seen for them in I1.}

\section{Evaluation}
\label{se:evaluation}

In our experiments we tackle the following research questions:
\begin{itemize}[topsep=1pt,itemsep=1pt]
  \item {\bf RQ 1.} How does \ourname compare to state-of-the-art fuzzers 
on targets that process chunk-based formats? 
  \item {\bf RQ 2.} Can \ourname identify new bugs?
  \item {\bf RQ 3.} How do tags relate to the actual input formats? And what is the role of structural mutations and roadblock bypassing in the observed improvements?
\end{itemize}

{\bf Benchmarks.} We consider the following programs (version, input format): {\sf wavpack} (5.1.0, WAV), {\sf decompress} (2.3.1, JP2), {\sf ffmpeg} (4.2.1, AVI), {\sf libpng} (1.6.37, PNG), {\sf readelf} (2.3.0, ELF), {\sf djpeg} (5db6a68, JPEG), {\sf objdump} (2.31.51, ELF), {\sf mpg321} (MP3, 0.3.2), {\sf oggdec} (1.4.0, OGG), {\sf tcpdump} (4.9.2, PCAP), and {\sf gif2rgb} (5.2.1, GIF). The first 6 programs were considered in past evaluation of \aflsmart, with a format specification available for it. The last 8 programs are commonly used in evaluations of general-purpose fuzzers.

\smallskip
{\bf Experimental setup.} We ran our tests on a server with two Intel Xeon E5-4610v2@2.30GHz CPUs and 256 GB of RAM, running Debian 9.2. We measured the cumulative basic block coverage of different fuzzers from running an application over the entire set of inputs generated by each fuzzer. We repeated each experiment 5 times, plotting the median value in the charts. For the second stage of \ourname we set Pr$_{field} =\ $Pr$_{chunk} = 1/15$ (Figure~\ref{fig:weizz-overview}), similarly to the probability of applying smart mutations in \aflsmart in~\cite{aflsmart-tse}. 



\begin{figure}[t]
  \centering
  \adjustbox{max width=0.79\columnwidth}{ 
  \input{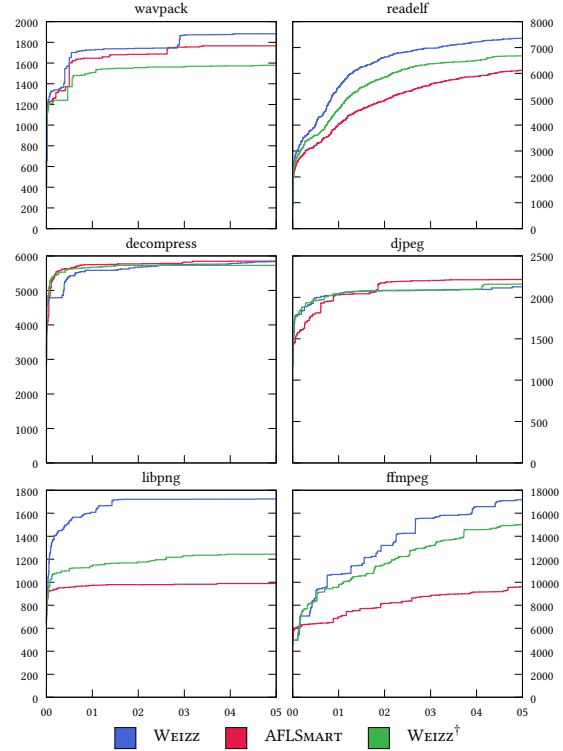}
  }
  \vspace{-3mm}
  \begin{table}[H]
  \centering
  \adjustbox{max width=0.6\columnwidth}{
    \footnotesize
    \centering
    \begin{tabular}{ l l l }
     \fcolorbox{black}[HTML]{4363D8}{\rule{0pt}{2pt}\rule{2pt}{0pt}}\quad \weizz & \fcolorbox{black}[HTML]{E6194B}{\rule{0pt}{2pt}\rule{2pt}{0pt}}\quad \aflsmart & \fcolorbox{black}[HTML]{3CB44B}{\rule{0pt}{2pt}\rule{2pt}{0pt}}\quad {\weizz}$^{\dagger}$ \\ 
    \end{tabular}
      }
  \end{table}
  \vspace{-7mm}
  \vspace{-1pt}
\caption{Basic block coverage over time (5 hours).\label{fig:comparison-aflsmart}}
\vspace{-1mm}
\end{figure}

\vspace{-1mm}
\subsection{RQ1: Chunk-Based Formats}
\label{ss:eval-coverage}

We compare \ourname against the state-of-the-art solution for chunk-based formats \aflsmart, which applies higher-order mutations over a virtual input structure (Section~\ref{ss:related-format}). We then take into account general-purpose fuzzers that previous studies~\cite{aflsmart-tse, ECLIPSER-ICSE19} suggest as being still quite effective in practice on this type of programs. 

\vspace{-1mm}
\subsubsection{{\scshape{AFLSmart}}} 
\label{ss:comparison-aflsmart}


For \aflsmart we used its release {\tt 604c40c} and the peach pits written by its authors for the virtual input structures involved. We measured the code coverage achieved by: (a) \aflsmart with stacked mutations but without deterministic stages as suggested in its documentation, (b) \ourname in its standard~\pagebreak[4] configuration, and (c) a variant {\ourname}$^{\dagger}$ with {\sc FuzzOperands}, checksum patching, and input repairing disabled. {\ourname}$^{\dagger}$ lets us focus on the effects of tag-based mutations, giving up on roadblock bypassing capabilities missing in \aflsmart. We provide (a) and (c) with a dictionary of tokens for format under analysis like in past evaluations of \aflsmart. We use \afl test cases as seeds, except for wavpack and ffmpeg where \okrev{we use minimal syntactically valid files}~\cite{small-files-project}.

\begin{figure}[t]
  \centering 
 \adjustbox{max width=0.79\columnwidth}{
  \input{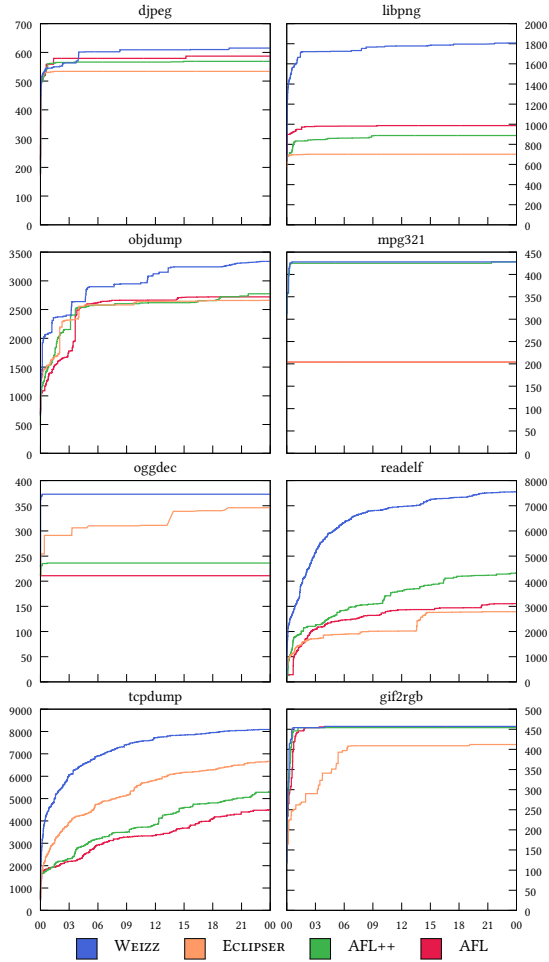}
  }
  \vspace{-3mm}
  \vspace{-1pt}
  \begin{table}[H]
  \centering
   \adjustbox{max width=0.7\columnwidth}{
    \footnotesize
    \centering
    \begin{tabular}{ c c c c }
     \fcolorbox{black}[HTML]{4363d8}{\rule{0pt}{2pt}\rule{2pt}{0pt}}\quad \weizz & \fcolorbox{black}[HTML]{FF9966}{\rule{0pt}{2pt}\rule{2pt}{0pt}}\quad \eclipser & \fcolorbox{black}[HTML]{3cb44b}{\rule{0pt}{2pt}\rule{2pt}{0pt}}\quad \aflpp & \fcolorbox{black}[HTML]{e6194b}{\rule{0pt}{2pt}\rule{2pt}{0pt}}\quad \afl
    \end{tabular}
  }
  \end{table}
  \vspace{-7mm}
   \vspace{-1pt}
  \caption{Basic block coverage over time (24 hours).\label{fig:comparison-other-fuzzers}}
  \vspace{-3mm}
\end{figure}



Figure~\ref{fig:comparison-aflsmart} plots the median basic block coverage after 5 hours. 
Compared to \aflsmart, \ourname brings \okrev{appreciably} higher coverage on 3 out of 6 subjects (readelf, libpng, ffmpeg), slightly higher on wavpack, comparable on decompress, and slightly worse on djpeg. To understand these results, we first consider where \weizzOne lies, and then discuss the benefits from the additional features of \ourname.



Higher coverage in \weizzOne comes from the different approach to structural mutations. \aflsmart follows a format specification, while we rely on how a program handles the input bytes: \weizzOne can reveal behaviors that characterize the actual implementation and that may not be necessarily anticipated by the specification. The first implication is that \weizzOne mutates only portions that the program has already processed in the execution, as tags derive from executed comparisons. The second is that imprecision~\pagebreak in inferring structural facts may actually benefit \weizzOne. The authors of \aflsmart acknowledge how relaxed specifications can expose imprecise implementations~\cite{aflsmart-tse}: we will return to this in Section~\ref{se:discussion}. 

\weizzOne is a better alternative than \aflsmart for readelf, libpng, and ffmpeg. When we consider the techniques disabled in \weizzOne, \ourname brings higher coverage for two reasons. I2S facts help \ourname replace magic bytes only in the right spots compared to dictionaries, which can also be incomplete. This turns out important also in multi-format applications like ffmpeg. Then, checksum bypassing allows it to generate valid inputs for programs that check data integrity, such as libpng that computes CRC-32 values over data.


We also consider the collected crashes, found only for wavpack and ffmpeg. In the first case, the three fuzzers found the same bugs. For ffmpeg, \ourname found a bug from a division by zero in a code section handling multiple formats: we reported the bug and its developers promptly fixed it. The I2S-related features of \ourname were very effective for generating inputs that deviate significantly from the initial seed, e.g., mutating an AVI file into an MPEG-4 one. In Section~\ref{se:real-world-bugs} \okrev{we back this claim with} one case study.




\vspace{-1mm}
\subsubsection{Grey-Box Fuzzers}
\label{ss:comparison-other-fuzzers}

We now compare \ourname against 8 popular applications handling chunk-based formats, heavily tested by the community~\cite{google-oss-fuzz}, and used in past evaluations of state-of-the-art fuzzers~\cite{vuzzer,Li:2017:SPB:3106237.3106295,redqueen,aflsmart-tse}.
%
%
We tested the following fuzzers: (a) \afl 2.53b in QEMU mode, (b) \aflpp 2.54c in QEMU mode enabling the CompareCoverage feature, and (c) \eclipser with its latest release when we ran our tests (commit {\tt 8a00591}) and its default configuration, aligned with \klee. Note that this is only one of the possible configurations of \eclipser, in particular this one is limited to a small file length.

As \ourname targets binary programs, we rule out fuzzers like \angora that require source instrumentation. For sub-instruction profiling, since \steelix is not publicly available, we choose \aflpp instead. 

While considering \redqueen could be \rev{tantalizing}, its hardware-assisted instrumentation could make an unfair edge as here we compare coverage of QEMU-based proposals, as the different overheads may mask why a given methodology reaches some coverage.  We attempt an experiment in Section~\ref{se:tag-impact} by configuring \ourname to resemble its features, and compare the two techniques  in Section~\ref{se:discussion}.



As the competitors we consider have no provisions for structural mutations, we opted for a larger budget of 24 hours in order to see whether they could recover in terms of coverage over time. Consistently with previous evaluations~\cite{ECLIPSER-ICSE19}, we use as initial seed a string made of the ASCII character "0" repeated 72 times. However, for libpng and tcpdump we fall back to a valid initial input test ({\tt not\_kitty.png} from \afl for libpng and a PCAP of a few seconds for tcpdump) as no fuzzer, excluding \ourname on libpng, achieved significant  coverage with artificial seeds. We also provide \afl with format-specific dictionaries to aid it with magic numbers.

Figure~\ref{fig:comparison-other-fuzzers} plots the median basic block coverage over time. \ourname on 5 out 8 targets (libpng, oggdec, tcpdump, objdump and readelf) achieves significantly higher code coverage than other fuzzers. The first three process formats with checksum fields that only \ourname repairs, although \eclipser appears to catch up over time for oggdec. Structural mutation capabilities combined with this factor may explain the gap between \ourname and other fuzzers. For objdump and readelf, we may speculate I2S facts are boosting the work of \ourname by the way \redqueen outperforms other fuzzers over them in its evaluation~\cite{redqueen} (in objdump logging function arguments was crucial~\cite{redqueen}). 
%
%
%
On mpg321 and gif2rgb, \afl-based fuzzers perform very similarly, with a large margin over \eclipser, confirming that standard \afl mutations can be effective \okrev{on some} chunk-based formats. Finally, \ourname leads for djpeg but the other fuzzers are not too far from it. 
Overall, \afl is interestingly the best alternative to \ourname for djpeg, libpng, and gif2rgb. When taking into account crashes, \ourname and \aflpp generated the same crashing inputs for mpg321, while only \ourname revealed a crash for objdump.



\subsection{RQ2: New Bugs}
\label{se:real-world-bugs}


To explore its effectiveness, we ran \ourname for 36 hours on several real-world targets, including the processing of inputs not strictly adhering to the chunk-based paradigm. \ourname revealed 16 bugs in 9 popular, well-tested applications: objdump, CUPS (2 bugs), libmirage (2), dmg2img (3), jbig2enc, mpg321, ffmpeg (3 in libavformat, 1 in libavcodec), sleuthkit, and libvmdk. Overall 6 bugs are NULL pointer dereferences (CWE-476), 1 involves an unaligned realloc (CWE-761), 2 can lead to buffer overflows (CWE-122), 2 cause an out-of-bounds read (CWE-125), 2 a division by zero (CWE-369), and 3 an integer overflow (CWE-190). We detail two interesting ones.



\smallskip
{\bf CUPS.} While the HTML interface of the Common UNIX Printing System is not chunk-oriented, we explored if \weizz could mutate the HTTP requests enclosing it. \ourname crafted a request that led CUPS to reallocate a user-controlled buffer, paving the road to a {\em House of Spirit} attack~\cite{house-of-spirit}. %
The key to finding the bug was to have {\sc FuzzOperands} \okrev{replace some current input bytes with I2S facts ({\tt 'Accept-Language'} logged as operand in a call to {\tt \_cups\_strcase} {\tt cmp}), thus materializing a valid request field that chunk mutations later duplicated}. Apple confirmed and fixed the bug in CUPS v2.3b8. 

\smallskip
{\bf libMirage.} We found a critical bug in a library providing uniform access to CD-ROM image formats. 
An attacker can generate a heap-based buffer overflow that in turn may corrupt allocator metadata, and even grant root access as the CDEmu daemon using it runs with root privileges on most Linux systems. We used an ISO image as initial seed: \ourname exposed the bug in the NRG image parser, demonstrating how it can deviate even considerably among input formats based exclusively on how the code handles input bytes.

\subsection{RQ3: Understanding the Impact of Tags}
\label{se:tag-impact}

Programs can differ significantly in how they process input bytes of different formats, yet we find it interesting to explore {\em why} \ourname can be effective on a given target. We discuss two case studies where we seek how tags can assist field and chunk identification in libpng, and we see how smart mutations and roadblocks bypassing are essential for ffmpeg, but either alone is not enough for efficacy.


\begin{figure}[t!]
\begin{center}
  \includegraphics[width=67mm]{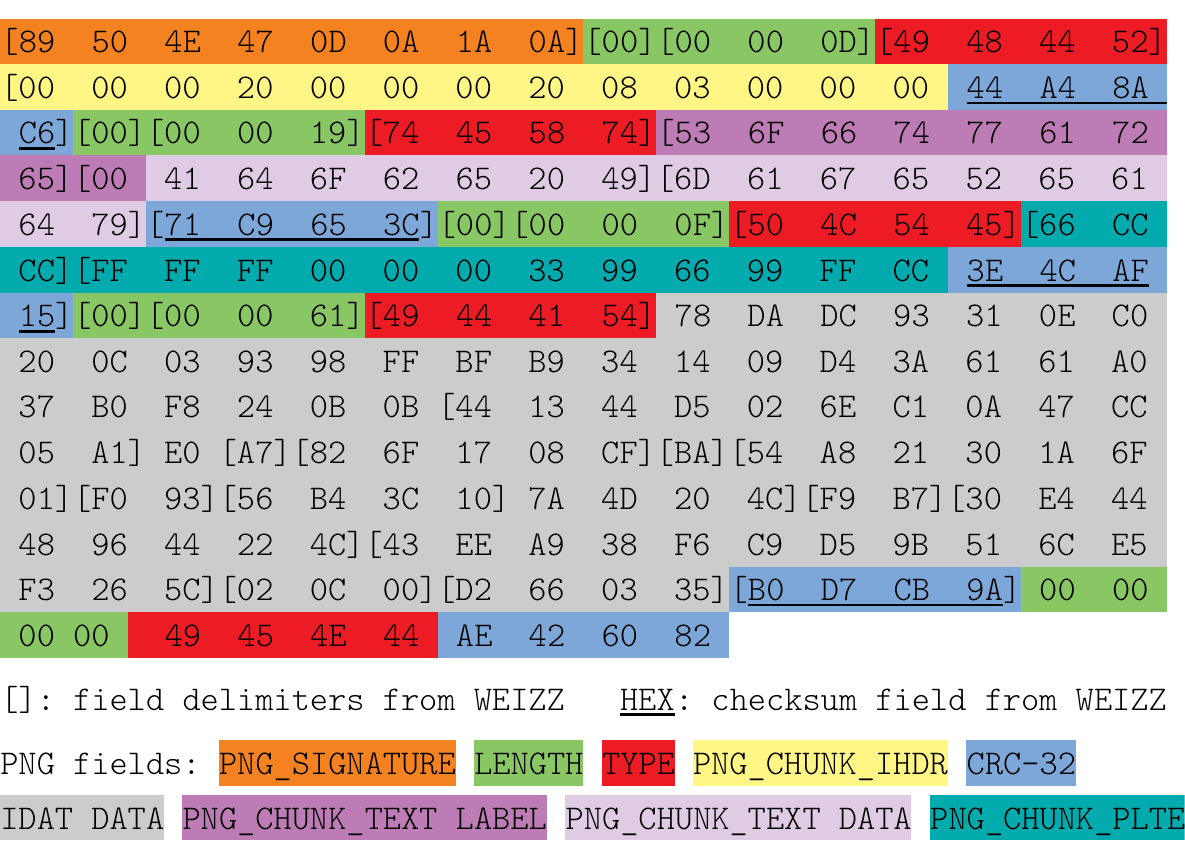}
 \end{center}
 \vspace{-4mm}
 \vspace{-1pt}
 \caption{Fields identified in the {\tt not\_kitty.png} test case.\label{fig:weizz-png}}
\end{figure}

\smallskip
{\bf Tag Accuracy.}
Figure~\ref{fig:weizz-png} shows the raw bytes of the seed {\tt not\_kit} {\tt ty.png} for libpng. It starts with a 8-byte magic number (in orange) followed by 5 chunks, each characterized by four fields: length (4 bytes, in green), type (4 bytes, red), data (heterogeneous, as with {\tt PNG\_CHUNK\_IHDR} data in yellow), and a CRC-32 checksum (4 bytes, in blue). The last chunk has no data as it marks the file end.


To understand the field identification process, we analyzed the tags from the surgical stage for the test case. Starting from the first byte, we apply {\sc FindFieldEnd} repeatedly at every position after the end of the last found field. In Figure~\ref{fig:weizz-png} we delimit each found field in square brackets, and underline bytes that \ourname deems from checksum fields. \ourname identifies correct boundaries with a few exceptions. The last three fields in the file are not revealed, as libpng never accesses that chunk. \okrev{In some cases a data field is merged with the adjacent checksum value. Initially libpng does not make comparisons over the data bytes, but first computes their checksum and verifies it against the expected value with a comparison that will characterize also the data bytes (for the data-flow dependency). Dependency analysis on the operands (Section~\ref{ss:checksumsAndFuzz}) however identifies the checksum correctly for {\sc FixChecksums}, and once the input gets repaired and enters the surgical stage again, libpng will execute {\em new} comparisons over the data bytes, and \ourname can identify the data field correctly tagging them with such instructions.} Finally, the grey parts ({\tt IDAT DATA}) represent a binary blob fragmented  into distinct fields by \ourname with gaps from untagged bytes, as not all the routines that process them make comparisons over the blob.

Analyzing chunk boundaries is more challenging, as {\sc FindFieldEnd} is sensitive to the initial position. For instance, when starting the analysis at the 8-byte magic number it correctly identifies the entire test case as one chunk, or by starting at a length field it may capture all the other fields for that chunk\footnotemark[3]. However, when starting at a type field it may build an incomplete chunk. Nonetheless, even under imprecise boundary detection \ourname can still make valuable mutations for two reasons. First, mutations that borrow bytes from other test cases check the same leading tag, and this yields for instance splicing of likely ``compatible'' incomplete chunks. Second, \okrev{this may be beneficial to exercise not so well-tested corner cases or shared parsing code}: we will return to it in Section~\ref{se:discussion}. For libpng we achieved better coverage than other fuzzers, including \aflsmart.

\footnotetext[3]{Depending on the starting byte, it could miss some of the first bytes.}

\begin{figure}[t]
  \centering
  
  \adjustbox{max width=0.65\columnwidth}{
  \input{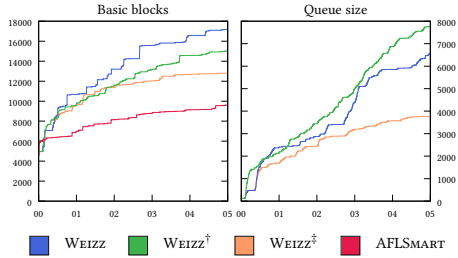}
  }
  \vspace{-2mm}
  \vspace{-1pt}
  \begin{table}[H]
    \footnotesize
    \centering
   \adjustbox{max width=0.7\columnwidth}{
   \centering
    \begin{tabular}{ c c c c }
     \fcolorbox{black}[HTML]{4363d8}{\rule{0pt}{2pt}\rule{2pt}{0pt}}\quad \weizz & \fcolorbox{black}[HTML]{3cb44b}{\rule{0pt}{2pt}\rule{2pt}{0pt}}\quad {\ourname}$^{\dagger}$ & \fcolorbox{black}[HTML]{FF9966}{\rule{0pt}{2pt}\rule{2pt}{0pt}}\quad {{\weizz}$^\ddagger$} & \fcolorbox{black}[HTML]{e6194b}{\rule{0pt}{2pt}\rule{2pt}{0pt}}\quad \aflsmart
    \end{tabular}
    }
  \end{table}
  \vspace{-5.5mm}
  \vspace{-1pt}
  \caption{Analysis of ffmpeg with three variants of \ourname.\label{fig:impact-tag-mutations}}
\end{figure}

\vspace{2pt}
{\bf Structural Mutations vs. Roadblocks.} To dig deeper in the experiments of Section~\ref{ss:eval-coverage}, one question could be: {\em how much coverage improvement is due to structural mutations or roadblock bypassing?} We consider ffmpeg as case study: in addition to \ourname and \weizzOne that lacks roadblock bypassing techniques but leverages structural mutations, we introduce a variant \weizzTwo that can use I2S facts to bypass roadblocks like \redqueen but lacks tag-based mutations. In Figure~\ref{fig:impact-tag-mutations} we report code coverage and input queue size after a run of 5 hours\footnotemark[4]. When taken individually, tag-based mutations seem to have an edge over roadblock bypass techniques (+17\% coverage). This may seem surprising as ffmpeg supports multiple formats and I2S facts can speed up the identification of their magic numbers, unlocking several program paths. Structural mutations however affect code coverage more appreciably on ffmpeg. When combining both features in \ourname, we get the best result with 34\% more coverage than in \weizzTwo. Interestingly, \weizzOne shows a larger queue size than \ourname: this means that although \ourname explores more code portions, \weizzOne covers some of the branches more exhaustively, i.e., it is able to cover more hit count buckets for some branches.

\footnotetext[4]{\aflsmart shown as reference---queues are uncomparable (no context sensitivity).}

\vspace{-1mm}
\section{Concluding Remarks}
\label{se:discussion}

\ourname introduces novel ideas for computing byte dependency information to simultaneously overcome roadblocks and back fully automatic fuzzing of chunk-based binary formats. The experimental results seem promising: we are competitive with human-assisted proposals, and we found new bugs in well-tested software.

Our approach has two practical advantages: fuzzers already attempt bitflips in deterministic stages, and instrumenting comparisons is becoming a common practice for roadblocks. We empower such analyses to better characterize the program behavior while fuzzing, enabling the tag assignment mechanism. Prior proposals do not offer sufficient information to this end: even for \redqueen, its colorization~\cite{redqueen} identifies I2S portions of an input (crucial for roadblocks) but cannot reveal dependencies for non-I2S bytes.

A downside is that bit flipping can get costly over large inputs. However, equally important is the time the program takes to execute one test case. In our experiments \ourname applied the surgical stage to inputs up to 3K bytes, with comparable or better coverage than the other fuzzers we considered. We leave to future work using forms of bit-level DTA~\cite{debray-scam14,debray-sp15} as an alternative for ``costly'' inputs.


\ourname may miss dependencies for comparisons made with uninstrumented instructions. This can happen in optimized code that uses arithmetic and logical operations to set CPU flags for a branch decision. We may resort to intra-procedural static analysis to spot them~\cite{vuzzer} (as logging all of them blindly can be expensive) but currently opted for tolerating some inconsistencies in the heuristics we use, for instance skipping over one byte in {\sc FindChunkEnd} when the remaining bytes would match the expected patterns.



Our structure-aware stage, in addition to not requiring a specification, is different than \aflsmart also in where we apply high-order operators. \aflsmart mutates chunks in a black-box fashion, that is, it has no evidence whether the program manipulated the involved input portion during the execution. \ourname chooses among chunks that the executed comparison instructions indirectly reveal. \rev{We find hard to argue which strategy is superior in the general case}. Another important difference is that, as our inference schemes are not sound, we may mutate inputs in odd ways, for instance replacing only portions of a chunk with a comparable counterpart from another input. \okrev{In the} \aflsmart paper the authors explain that they could find some bugs only \okrev{thanks to} a relaxed specification~\cite{aflsmart-tse}. We find this consistent with the \grimoire experience with grammars, where paths outside the specification revealed coding mistakes~\cite{grimoire}.

As future work we plan to consider a larger pool of subjects, and shed more light on the impact of structure-aware techniques: how they impact coverage, which are the most effective for a program, and how often the mutated inputs do not meet the specification. Answering these questions seems far from trivial due to the high throughput and entropy of fuzzing. There is also room to extend chunk inference with new heuristics or make the current ones more efficient. For instance, we are exploring with promising results a variant where we locate the beginning of a chunk at a field made of I2S bytes, possibly indicative of a magic value for its type. 
\unless\ifdefined\daio
\balance
\fi
\balance
\bibliographystyle{ACM-Reference-Format}
\bibliography{bibliography}

\ifdefined\daio
\clearpage
\nobalance
\appendix

\renewcommand{\thesection}{\Alph{section}}
\setcounter{section}{0}
\section{Appendix}
\label{se:apx}

\subsection{Roadblocks from lodepng}
\label{apx:roadblocks}

As an example of roadblocks found in real-world software, the following excerpt from {\tt lodepng} image encoder and decoder looks for the presence of the magic string {\tt IHDIR} denoting the namesake PNG chunk type and checks the integrity of its data against a CRC-32 checksum:

\vspace{-1mm}
\begin{scozzo}
if (!lodepng_chunk_type_equals(in + 8, "IHDR")) {
  /*error: it doesn't start with a IHDR chunk!*/
  CERROR_RETURN_ERROR(state->error, 29);
}
...
unsigned CRC = lodepng_read32bitInt(&in[29]);
unsigned checksum = lodepng_crc32(&in[12], 17);
if (CRC != checksum) {
  /*invalid CRC*/
  CERROR_RETURN_ERROR(state->error, 57);
}
\end{scozzo}
\vspace{-4mm}

\subsection{Additional Algorithms}

Algorithm~\ref{alg:weizz-phase-one} formalizes the workflow of the surgical stage depicted in Figure~\ref{fig:weizz-overview}. With Figure~\ref{fig:glossary} we then provide our readers with a glossary of the domain and data structure definitions used in the algorithmic descriptions of the techniques we propose.

\subsubsection*{Tag Placement} Algorithm~\ref{alg:weizz-tags} provides pseudocode for the {\sc PlaceTags} procedure, while we refer the reader to Section~\ref{ss:tag-placement} for a discussion of its main steps. An interesting detail presented here is related to the condition $n'>4$ when computing the {\em reassign} flag. This results from experimental observations, with 4 being the most common field size in the formats we analyzed. Hence, \ourname will choose as characterizing instruction for a byte the first comparison that depends on no more than 4 bytes, ruling out possibly ``shorter'' later comparisons as they may be accidental.


\subsubsection*{Checksum Handling}
Due to space limitations in the paper we have provided our readers only with the intuition of how our checksum handling schemes works. We detail its main steps in the following.

As we said the last conceptual step of the surgical stage involves checksum validation and input repair with respect to checksums patched in the program \okrev{during previous surgical iterations}. In the process \ourname also discards false positives among such checksums.

{\sc FixChecksums} (Algorithm~\ref{alg:weizz-fix-checksum}) filters tags marked as involved in checksum fields \okrev{for which a patch is active, and processes them in the order described in the previous section}. \ourname initially executes the patched program, obtaining a comparison table CT and \okrev{the hash $h$ of the coverage map as execution path footprint}.

Then it iterates over each filtered tag as follows. It first extracts the input-derived operand value of the comparison instruction (i.e., the checksum computed by the program) and where in the input the contents of the other operand (which is I2S, see Section~\ref{ss:checksumsAndFuzz}) are located, then it replaces the input bytes with the expected value (lines 4-5). \ourname makes a new instrumented execution over the repaired input, disabling the patch. If it handled the checksum correctly, the resulting hash $h^{\prime}$ will be identical to $h$ implying no execution divergence happened, \okrev{otherwise the checksum is a false positive and we have to bail}.

After processing all the tags, we may still have patches left in the program for checksums not reflected by tags. \ourname may in fact erroneously treat a comparison instruction as a checksum due to inaccurate dependency information. \okrev{\ourname removes all such patches and if after that the footprint for the execution does not change, the input repair process succeeded. This happens typically with patches for branch outcomes that turn out to be already ``implied'' by the repaired tag-dependent checksums.}
Otherwise \ourname tries to determine \okrev{which patch not reflected by tags may individually cause a divergence: it tries patches one at a time and compares footprints to spot offending ones}. This happens when \ourname initially finds a checksum for an input and the analysis of a different input reveals an overfitting, so we mark it as a false positive.


\begin{figure}[t]
\centering
\begin{footnotesize}
\adjustbox{max width=0.85\columnwidth}{
\begin{tabular}{|l|p{4.2cm}|}
\hline
$J$: $\{0\,...\,2^{8}-1\}$ & Index of {cmp} instance \\
$Sites$: $Addr\times Stack$ $\to$ $\{0\,...\,2^{16}$-$1\}$ & Site ID (cmp addr$\,\oplus\,$calling context) \\ 
$CT$: $Sites\times J \times \{\textit{op}_1\textit{, op}_2\}$ $\to$ $V$ & cmp site instance \& operand$\,\to\,$value \\
$Deps$: $Sites\times J \times \{\textit{op}_1\textit{, op}_2\}$ $\to$ $A$ & cmp site instance \& operand$\,\to$ array of len(input) booleans \\
$R$: $Sites\times J$ $\to$ $InputToStateInfo$ & cmp site instance$\,\to$ I2S info \\
$CI$ : $Addr \to ChecksumInfo$ & cmp addr$\,\to$ checksum info \\
\hline
\end{tabular}
}
\end{footnotesize}
\vspace{-3.5mm}
\caption{Glossary and domain definitions.\label{fig:glossary}}
\vspace{-2mm}
\end{figure}

\begin{figure}[t]
\vspace{-1mm}
\removelatexerror
\setlength{\textfloatsep}{0mm}
\begin{algorithm}[H]
\footnotesize
  \DontPrintSemicolon
  \SetAlgoNoEnd
  \SetAlgoNoLine
  \SetNlSkip{-0.5em}
  \SetKwFunction{FMain}{\sc SurgicalFuzzing}
  \SetKwProg{Fn}{function}{:}{}
  \Fn{\FMain{I}}{
  \Indmm
    \nextnr CT, Deps $\gets$ {\sc GetDeps}(I)\;
      \nextnr \ForEach{s $\in$ Sites, j $\in$ J}{\Indmm
         \nextnr R(s, j) $\gets$ {\sc DetectI2S}(Deps, CT, s, j, I)\;
         \nextnr CI $\gets$ {\sc MarkChecksums}(R(s, j), CI)\;
         \nextnr {\sc FuzzOperands}(I, CI, Deps, s, j)\;
       }
    \nextnr Tags $\gets \langle0\,...\,0\rangle$\ with $|$Tags$|$ = len(I)\;
    \nextnr o $\gets$ {\sc TopologicalSort}(CI, Deps)\;
    \nextnr \ForEach{s: {\sc SortSitesByExecOrder}(CT)}{\Indmm
      \nextnr Tags $\gets$ {\sc PlaceTags}(Tags, Deps, s, CT, CI, R, o, I)\;
    }
    \nextnr I, CI, fixOk $\gets$ {\sc FixChecksums(CI, I, Tags, o)}\;
    \nextnr CI $\gets$ {\sc PatchAllChecksums(CI)}\;
    \nextnr \Return{I, Tags, fixOk}\;
  }
  \vspace{2pt}
  \caption{\Firststage stage of \ourname.\label{alg:weizz-phase-one}}
\end{algorithm}
\setlength{\textfloatsep}{\textfloatsepsave}
\vspace{-2mm}
\end{figure}

\begin{figure}[t!]
\removelatexerror
\begin{algorithm}[H]
\footnotesize
  \DontPrintSemicolon
  \SetAlgoNoEnd
  \SetAlgoNoLine
  \setstretch{1.1} 
  \SetNlSkip{-0.45em}
  \SetKwFunction{FMain}{\sc PlaceTags}
  \SetKwProg{Fn}{function}{:}{}
  \Fn{\FMain{Tags, Deps, s, CT, CI, R, o, I}}{
  \Indmm
    \nextnr \For{b$\,\in\,$\{0\,...\,len(I)-1\}, op$\,\in\{\textit{op}_1\textit{, op}_2\}$}{\Indmm\Indmm
        \nextnr \lIf{$\neg\bigvee_j Deps(s, j, op)[b]$}{{\bf continue}} 
        \nextnr n $\gets \sum_k  (1~\textbf{if}\,\bigvee_j Deps(s, j, op)[k]$~\textbf{else}~0)\;
        \nextnr n$^{\prime}$ $\gets$ Tags[b].numDeps\;
        \nextnr reassign $\gets$ (!{\sc IsCksm}(Tags[b]) $\wedge$ n$^{\prime}$ > 4 $\wedge$ n < n$^{\prime}$)\;
        \nextnr \If{b $\not\in$ Tags$\,\vee\,${\sc IsValidCksm}(CI, s)$\,\vee\,$reassign}{\Indmm
          \nextnr Tags[b] $\gets$ {\sc SetTag}(Tags[b], CT, s, op, CI, R, n, o)\;
      }
    }
    \nextnr \Return{Tags}\;
  }
  \vspace{2pt}
  \caption{{\sc PlaceTags} procedure.\label{alg:weizz-tags}}
\end{algorithm}
\end{figure}

\begin{figure}[t!]
\vspace{-2mm}
\removelatexerror
\begin{algorithm}[H]
\footnotesize
  \DontPrintSemicolon
  \SetAlgoNoEnd
  \SetAlgoNoLine
  \SetNlSkip{-0.45em}
  \SetKwFunction{FMain}{\sc FixChecksums}
  \SetKwProg{Fn}{function}{:}{}
  \Fn{\FMain{CI, I, Tags, o}}{
  \Indmm
    \nextnr Tags$_{ck}$ $\gets$ {\sc FilterAndSortTags}(Tags, o)\;
    \nextnr CT, h $\gets$ {\sc RunInstr}$_{ck}$(I)\;
    \nextnr \For{k $\in$ \{0\,...\,$|$Tags$_{ck}$$|$-1\} }{\Indmm
      \nextnr v, idx $\gets$ {\sc GetCmpOpOfTag}(CT, Tags$_{ck}$[k], CI, R)\;
      \nextnr I $\gets$ {\sc FixInput}(I, v, idx)\;
      \nextnr {\sc UnpatchProgram}(Tags$_{ck}$[k], CI)\;
      \nextnr CT, h$^{\prime}$ $\gets$ {\sc RunInstr}$_{ck}$(I$^{\prime}$)\;
      \nextnr \If{h $\neq$ h$^{\prime}$}{
        \nextnr CI $\gets$ {\sc MarkCksmFP}(CI, Tags$_{ck}$[k].id)\;
        \nextnr \Return{CI, I, false}\;
      }
    }
    \nextnr {\sc UnpatchUntaggedChecksums}(CI, Tags$_{ck}$)\;
    \nextnr h$^{\prime}$ $\gets$ {\sc RunLightInstr}$_{ck}$(I)\;
    \nextnr \lIf{h$^{\prime}$ == h}{\Return{CI, I, true}}
      \nextnr \ForEach{a $\in$ dom(CI) $|$ $\nexists$ t: Tags$_{ck}$[t].id == a}{\Indmm
        \nextnr {\sc PatchProgram}(C, a)\;
        \nextnr h$^{\prime\prime}$ $\gets$ {\sc RunLightInstr}$_{ck}$(I)\;
        \nextnr  \lIf{h$^{\prime\prime}$ $\neq$ h$^{\prime}$}{
          CI $\gets$ {\sc MarkCksmFP}(CI, a)
        }
        \nextnr {\sc UnpatchProgram}(C, a)\;
      }
    \nextnr \Return{CI, I, false}\;
  }
  \vspace{1mm}
  \caption{{\sc FixChecksums} procedure.\label{alg:weizz-fix-checksum}}
\end{algorithm}
\vspace{-3mm}
\end{figure}

Finally, as we explained in the paper \ourname at the end of the process re-applies patches that were not a false positive, so to ease the second stage and future surgical iterations over other inputs. \ourname also applies patches for checksums newly discovered by {\sc MarkChecksums} in the current surgical stage. 



%
\begin{figure}[t]
\removelatexerror
\begin{algorithm}[H]
\footnotesize
  \DontPrintSemicolon
  \SetAlgoNoEnd
  \SetAlgoNoLine
  \SetNlSkip{-0.45em}
  \SetKwFunction{FMain}{\sc FindFieldEnd}
  \SetKwProg{Fn}{function}{:}{}
  \Indmm
  \Fn{\FMain{Tags, k, depth}}{
    \nextnr end $\gets$ {\sc GoRightWhileSameTag}(Tags, k)\;
    \nextnr \If{depth $<$ 8 $\wedge$ Tags[end+1].ts == Tags[k].ts+1}{\Indmm
      \nextnr end $\gets$ {\sc FindFieldEnd}(Tags, end+1, depth+1)\;
    }
    \nextnr \Return{end}\;
  }
  \vspace{1mm}
  \caption{{\sc FindFieldEnd} procedure.\label{alg:weizz-field-end}}
\end{algorithm}
\vspace{-4mm}
\end{figure}


\begin{figure}[t]
\removelatexerror
\begin{algorithm}[H]
\footnotesize
  \DontPrintSemicolon
  \SetAlgoNoEnd
  \SetAlgoNoLine
  \SetNlSkip{-0.45em}
  \SetKwFunction{FMain}{\sc GetRandomChunk}
  \SetKwProg{Fn}{function}{:}{}
   \SetKwIF{With}{ElseWith}{}{with}{do}{else with}{}{endwith}
  \SetKwRepeat{Do}{do}{while}
  \Fn{\FMain{Tags}}{
    \nextnr \With{probability Pr$_{{chunk}_{12}}$}{
      \nextnr k $\gets$ {\sc Rand}(len(Tags))\;
      \nextnr start $\gets$ {\sc GoLeftWhileSameTag}(Tags, k)\;
      \nextnr end $\gets$ {\sc FindChunkEnd}(Tags, k, 0)\;
      \nextnr \Return{(start, end)}\;
    }
    \nextnr \Else{
      \nextnr id $\gets$ pick u.a.r. from $\bigcup_i$\{Tags[i].id\}\;
      \nextnr start $\gets$ 0, L $\gets$ $\emptyset$\;
      \nextnr \While{start < $|$Tags$|$}{
          \nextnr start $\gets$ {\sc SearchTagRight}(Tags, start, id)\;
          \nextnr end $\gets$ {\sc FindChunkEnd}(Tags, start)\;
          \nextnr L $\gets$ L $\cup$ \{(start, end)\}, start $\gets$ end + 1\;
       }
      \nextnr \Return{l $\in$ L chosen u.a.r.}\;
    }
  }
  \vspace{1mm}
  \caption{{\sc GetRandomChunk} procedure.\label{alg:weizz-chunk}}
\end{algorithm}
\vspace{-2mm}
\end{figure}

\subsubsection*{Field Identification} Algorithm~\ref{alg:weizz-field-end} shows how to implement the {\sc FindFieldEnd} procedure that we use to detect fields complying to one of the patterns from Figure~\ref{fig:field}, which we discussed in Section~\ref{ss:fields-mutations}. To limit the recursion at line 3 for field extension, we use 8 as maximum depth (line 7).

\subsubsection*{Chunk Identification} As discussed in Section~\ref{ss:chunk-mutations}, in some formats there might be dominant chunk types, and choosing the initial position randomly would reveal a chunk of non-dominant type with a smaller probability. \ourname chooses between the original and the alternative heuristic with a probability Pr$_{{chunk}_{12}}$ as described in Algorithm~\ref{alg:weizz-chunk}.


\subsection{Implementation Details}

Since the analysis of branches for \ourname is context-sensitive and the number of contexts in programs is often large~\cite{hcct-spe16}, we extend like in \angora the number of buckets in the coverage map from $2^{16}$ as in \afl to $2^{18}$, and compute the index using the source and destination basic block addresses and a one-word hash of the call stack. For $CT$ we choose to host up to $2^{16}$ entries since there is one less degree of freedom in computing indexes in the $Sites$ function (Figure~\ref{fig:glossary}). In the remainder of the section we present present two optimizations that we integrated in our implementation. 

\subsubsection*{Loop Bucketization} 
In addition to newly encountered edges, \afl deems a path interesting when some branch hit count changes (Section~\ref{ss:related-cgf}). To avoid keeping too many inputs, \afl verifies whether the counter falls in a previously unseen power-of-two interval. However, as pointed out by the authors of \lafintel~\cite{lafintel}, this approach falls short in deeming progress with patterns like the following:
\begin{scozzo}
for (int i = 0; i < strlen(magic); ++i)
    if (magic[i] != input[i]) return;
\end{scozzo}

\noindent
To mitigate this problem, in the surgical phase \ourname~\pagebreak considers interesting also inputs that lead to the largest hit count observed for an edge  across the entire stage. 

\subsubsection*{Optimizing I2S Fuzzing} 
Another tuning for the surgical stage involves the {\sc FuzzOperands} step. \ourname decides to skip a comparison site when previous iterations over it were unable to generate any coverage-improving paths. In particular, the probability of executing {\sc FuzzOperands} on a site is initially one and decreases with the ratio between failed and total attempts.

\subsection{Additional Experimental Data}
\label{appendix-eval}

In the following we provide two tables that recap the benchmarks used in the evaluation (Table~\ref{tab:targets}) and provide references for the bugs we found in the programs mentioned in Section~\ref{se:real-world-bugs} (Table~\ref{tab:bugs}), respectively. We then provide two tables where we show the 60\%-confidence interval for the median value in the charts of Figure~\ref{fig:comparison-aflsmart} (Table~\ref{tab:comparison-aflsmart-ci}) and Figure~\ref{fig:comparison-other-fuzzers} (Table~\ref{tab:comparison-other-fuzzers-ci}). Finally, we present two tables that summarize the best alternative to \ourname for each benchmark considered in Figure~\ref{fig:comparison-aflsmart} (Table~\ref{tab:comparison-best-alternative}) and Figure~\ref{fig:comparison-other-fuzzers} (Table~\ref{tab:comparison-best-alternative}), and details on the crashes found by the fuzzers.

\begin{table}[ht]
  \caption{Target applications considered in the evaluation.\label{tab:targets}
  }
  \vspace{-3.5mm}
  \begin{adjustbox}{width=0.725\columnwidth,center}
  \small
  \begin{tabular}{ |l|l|l| }
    \hline
    {\sc Program} & {\sc Release} & {\sc Input Format} \\
    \hline
    wavpack & {\tt 5.1.0} & WAV \\
    decompress (openjp2) & {\tt 2.3.1-git} & JP2 \\
    ffmpeg & {\tt 4.2.1} & AVI \\
    \hline
    libpng & {\tt 1.6.37} & PNG \\ 
    readelf (GNU binutils) &  {\tt 2.3.0} &  ELF  \\
    djpeg (libjpeg-turbo) & {\tt git:5db6a68} & JPEG \\
    \hline
    objdump (GNU binutils) & {\tt 2.32.51} & ELF \\
    mpg321  & {\tt 0.3.2} & MP3 \\
    oggdec (vorbis-tools) & {\tt 1.4.0} & OGG  \\
    tcpdump & {\tt 4.9.2} & PCAP \\ 
    gif2rgb (GIFLIB) & {\tt 5.2.1} &  GIF \\
    \hline
  \end{tabular}
  \end{adjustbox}
  \vspace{-2mm}
\end{table}

\begin{table}[hb]
  \caption{Bugs found and reported in real-world software.\label{tab:bugs}}
  \vspace{-3.5mm}
  \centering
  \begin{adjustbox}{width=0.725\columnwidth,center}
  \small
  \begin{tabular}{ |l|l|l| }
    \hline
    Program & Bug ID & Type \\
    \hline
    objdump & Bugzilla \#24938 & CWE-476\\
    CUPS & rdar://problem/50000749 & CWE-761\\
    CUPS & GitHub \#5598 & CWE-476\\
    libmirage (CDEmu) & CVE-2019-15540 & CWE-122\\
    libmirage (CDEmu) & CVE-2019-15757 & CWE-476\\
    dmg2img & Launchpad \#1835461 & CWE-476\\
    dmg2img & Launchpad \#1835463 & CWE-125\\
    dmg2img & Launchpad \#1835465 & CWE-476\\
    jbig2enc & GitHub \#65 & CWE-476\\
    mpg321 & Launchpad \#1842445 & CWE-122\\
    libavformat (FFmpeg) & Ticket \#8335 & CWE-369\\
    libavformat (FFmpeg) & Ticket \#8483 & CWE-190\\
    libavformat (FFmpeg) & Ticket \#8486 & CWE-190\\
    libavcodec (FFmpeg) & Ticket \#8494 & CWE-190\\
    libvmdk & GitHub \#22 & CWE-369\\
    sleuthkit & GitHub \# 1796 & CWE-125\\
    \hline
  \end{tabular}
  \end{adjustbox}
  \vspace{-3mm}
\end{table}

\begin{table}[b]
  \centering
  \caption{Basic block coverage (60\% confidence intervals) after 5-hour fuzzing of the top 6 subjects of Table~\ref{tab:targets}.\label{tab:comparison-aflsmart-ci}}
  \vspace{-3.5mm}
  \begin{adjustbox}{width=0.7\columnwidth,center}
  \small
  \begin{tabular}{ |l||c|c|c| }
    \hline
    {\sc Programs} & \weizz & \aflsmart & {\ourname}$^{\dagger}$ \\
    \hline
    wavpack & {\bf 1824-1887} & 1738-1813 & 1614-1749 \\
    readelf & {\bf 7298-7370} & 6087-6188 & 6586-6731 \\
    decompress & 5831-6276 & {\bf 6027-6569} & 5376-5685 \\
    djpeg & 2109-2137 & {\bf 2214-2221} & 2121-2169 \\
    libpng & {\bf 1620-1688} & 1000-1035 & 1188-1231 \\
    ffmpeg & {\bf 15946-17885} & 9352-9923 & 14515-14885 \\
    \hline
  \end{tabular}
  \end{adjustbox}
\end{table}

\clearpage

\begin{table}[t]
  \centering
  \caption{Basic block coverage (60\% confidence intervals) after 24-hour fuzzing of the bottom 8 subjects of Table~\ref{tab:targets}.\label{tab:comparison-other-fuzzers-ci}}
  \vspace{-3.5mm}
  \begin{adjustbox}{width=0.75\columnwidth,center}
  \small
  \begin{tabular}{ |l||c|c|c|c| }
    \hline
    {\sc Programs} & \weizz & \eclipser & \aflpp & \afl \\
    \hline
    djpeg & {\bf 612-614} & 492-532 & 561-577 & 581-592 \\
    libpng & {\bf 1747-1804} & 704-711 & 877-901 & 987-989 \\
    objdump & {\bf 3366-4235} & 2549-2648 & 2756-3748 & 2451-2723 \\
    mpg321 & {\bf 428-451} & 204-204 & 426-427 & 204-204 \\
    oggdec & {\bf 369-372} & 332-346 & 236-244 & 211-211 \\
    readelf & {\bf 7428-7603} & 2542-2871 & 4265-5424 & 2982-3091 \\
    tcpdump & {\bf 7662-7833} & 6591-6720 & 5033-5453 & 4471-4576 \\
    gif2rgb & 453-464 & 357-407 & 451-454 & {\bf 457-465} \\
    \hline
  \end{tabular}
  \end{adjustbox}
\end{table}

\begin{table}[ht]
  \centering
  \caption{\ourname vs. best alternative for targets of Figure~\ref{fig:comparison-aflsmart}: a basic block (BB) coverage percentage higher than 100 means that \ourname outperforms the best alternative fuzzer.\label{tab:comparison-best-alternative}}
  \vspace{-3.5mm}
  \begin{adjustbox}{width=0.8\columnwidth,center}
  \small
  \begin{tabular}{ |l|c|c|c| }
    \hline
  \thead{\sc Program} & \thead{\sc Best Alternative} & \thead{\sc BB \% w.r.t. Best} & \thead{\sc Crashes found by}\\
    \hline
    wavpack & \aflsmart & 106\% & all\\
    readelf & {\ourname}$^{\dagger}$ & 110\% & none \\
    decompress & \aflsmart & 100\% & none \\
    djpeg & \aflsmart & 96\% & none \\
    libpng & {\ourname}$^{\dagger}$ & 132\% & none \\
    FFmpeg & {\ourname}$^{\dagger}$ & 114\% & \weizz \\
    \hline
  \end{tabular}
  \end{adjustbox}
  \vspace{-1mm}
\end{table}

\begin{table}[ht]
  \centering
  \caption{\ourname vs. best alternative for targets of Figure~\ref{fig:comparison-other-fuzzers}. Columns have the same meaning used in Table~\ref{tab:comparison-best-alternative}.\label{tab:comparison-fuzzers-best-alternative}}
  \vspace{-3.5mm}
  \begin{adjustbox}{width=0.8\columnwidth,center}
  \small
  \begin{tabular}{ |l|c|c|c| }
    \hline
  \thead{\sc Program} & \thead{\sc Best Alternative} & \thead{\sc BB \% w.r.t. Best} & \thead{\sc Crashes found by}\\
    \hline
    djpeg & \afl & 105\% & none \\
    libpng & \afl & 180\% & none \\
    objdump & \aflpp & 120\% & \weizz \\
    mpg321 & \aflpp & 100\% & \weizz, \aflpp\\
    oggdec & \eclipser & 108\% & none\\
    readelf & \aflpp & 175\% & none\\
    tcpdump & \eclipser & 117\% & none\\
    gif2rgb & \afl & 100\% & none\\
    \hline
  \end{tabular}
  \end{adjustbox}
  \vspace{-1mm}
  \vspace{-1pt}
\end{table}


\fi

\end{document}